\documentclass[AMA,Times2COL]{styles/nmr/WileyNJDv5} %STIX1COL,STIX2COL,STIXSMALL
\articletype{Article Type}%

\usepackage{styles/defs}
\usepackage{placeins}

\received{xx xxxxx xxxx}
\revised{xx xxxxx xxxx}
\accepted{xx xxxxx xxxx}
\begin{document}

% \jnlcitation{\cname{%
% \author{Williams K.}, 
% \author{B. Hoskins}, 
% \author{R. Lee}, 
% \author{G. Masato}, and 
% \author{T. Woollings}} (\cyear{2016}), 
% \ctitle{A regime analysis of Atlantic winter jet variability applied to evaluate HadGEM3-GC2}, \cjournal{Magn. Reson. Med.}, \cvol{2017;00:1--6}.}

\title{WAND: Wavelet Analysis-based Neural Decomposition of MRS Signals for Artifact Removal}

\author[1]{Julian P. Merkofer}%{\orcid{0000-0003-2924-5055}}
\author[2]{Dennis M. J. van de Sande}%{\orcid{0000-0001-6112-1437}}
\author[2]{Sina Amirrajab}%{\orcid{0000-0001-8226-7777}}
\author[3]{Kyung Min Nam}%{}
\author[1]{Ruud J. G. van Sloun}%{\orcid{0000-0003-2845-0495}}
\author[3]{Alex A. Bhogal}%{\orcid{0000-0003-3211-1760}}

\authormark{MERKOFER \textsc{et al}}
\titlemark{\MakeUppercase{WAND: Wavelet Analysis-based Neural Decomposition of MRS Signals for Artifact Removal}}

\address[1]{\orgdiv{Department of Electrical Engineering}, \orgname{Eindhoven University of Technology}, \orgaddress{\state{Eindhoven}, \country{The Netherlands}}}

\address[2]{\orgdiv{Department of Biomedical Engineering}, \orgname{Eindhoven University of Technology}, \orgaddress{\state{Eindhoven}, \country{The Netherlands}}}

\address[3]{\orgdiv{High Field Research Group, Center for Image Sciences},  \orgname{University Medical Center Utrecht}, \orgaddress{\state{Utrecht}, \country{The Netherlands}}}

\corres{Julian P. Merkofer, Eindhoven University of Technology , PO Box 513 5600 MB Eindhoven, The Netherlands. \email{\textcolor{blue}{j.p.merkofer@tue.nl}}}

\presentaddress{Eindhoven University of Technology
{\hfill\break}Electrical Engineering Department
{\hfill\break}Groene Loper 19, 5612 AP Eindhoven, The Netherlands}

\fundingInfo{project Spectralligence (EUREKA IA Call, ITEA4 project 20209) and NWO VIDI (VI.Vidi.223.085).}
% \JELinfo{ejlje}

% \finfo{This work was in part funded by Spectralligence (EUREKA IA Call, ITEA4 project 20209).}

%%%%%%%%%%%%%%%%
%%% Abstract %%%
%%%%%%%%%%%%%%%%
% \abstract[Summary]{
% % quick summary
% \section{Purpose}
% \section{Methods}
% \section{Results}
% \section{Conclusion}
% }
\abstract[Abstract]{
% 196/300 words...
Accurate quantification of metabolites in \ac{mrs} is challenged by low \ac{snr}, overlapping metabolites, and various artifacts. Particularly, unknown and unparameterized baseline effects obscure the quantification of low-concentration metabolites, limiting \ac{mrs} reliability. This paper introduces \ac{wand}, a novel data-driven method designed to decompose MRS signals into their constituent components: metabolite-specific signals, baseline, and artifacts. \Ac{wand} takes advantage of the enhanced separability of these components within the wavelet domain. The method employs a neural network, specifically a U-Net architecture, trained to predict masks for wavelet coefficients obtained through the continuous wavelet transform. These masks effectively isolate desired signal components in the wavelet domain, which are then inverse-transformed to obtain separated signals. Notably, an artifact mask is created by inverting the sum of all known signal masks, enabling \ac{wand} to capture and remove even unpredictable artifacts. The effectiveness of \ac{wand} in achieving accurate decomposition is demonstrated through numerical evaluations using simulated spectra. Furthermore, \ac{wand}'s artifact removal capabilities significantly enhance the quantification accuracy of linear combination model fitting. The method's robustness is further validated using data from the 2016 MRS Fitting Challenge and in-vivo experiments.
}

\keywords{magnetic resonance spectroscopy, wavelet analysis, deep learning, artifact removal}

% \jnlcitation{\cname{%
% \author{Williams K.}, 
% \author{B. Hoskins}, 
% \author{R. Lee}, 
% \author{G. Masato}, and 
% \author{T. Woollings}} (\cyear{2016}), 
% \ctitle{A regime analysis of Atlantic winter jet variability applied to evaluate HadGEM3-GC2}, \cjournal{Magn. Reson. Med.}, \cvol{2017;00:1--6}.}

\maketitle

%\footnotetext{\textbf{Abbreviations:}~\hbox{ANA,~anti-nuclear~antibodies;~APC,~antigen-}{\hfill\break}presenting~cells; IRF, interferon regulatory factor}

\footnotetext{\textbf{Abbreviations:} \acs{cnn},~\acl{cnn};~\acs{cwt},~\acl{cwt};~\acs{crlbp},~\acl{crlbp};~\acs{fmrs},~functional magnetic resonance spectroscopy;~\acs{hsvd},~\acl{hsvd};~\acs{icwt},~\acl{icwt};~\acs{lcm},~\acl{lcm};~\acs{mae},~\acl{mae};~\acs{mm},~\acl{mm};~\acs{mrs},~\acl{mrs};~\acs{mrsinmrs},~\acl{mrsinmrs};~\acs{mse},~\acl{mse};~\acs{nsa},~\acl{nsa};~\acs{press},~\acl{press};~\acs{se},~\acl{se};~\acs{sift},~\acl{sift};~\acs{snr},~\acl{snr};~\acs{wand},~\acl{wand};}

%%%%%%%%%%%%%%%%%%%%
%%% Introduction %%%
%%%%%%%%%%%%%%%%%%%%
\section{Introduction}\label{sec:intro}
\acresetall

Nuclear \ac{mrs} allows non-invasive determination of metabolite concentrations in biochemical samples of interest, making it a promising tool for diagnosing and monitoring various diseases, including cancer, neurological disorders, as well as traumatic brain injuries. \cite{faghihi_magnetic_2017, Horska2023MRSClincalA, Posse2013MRSI, Maudsley2020AdvancedMR} However, clinical adoption is hindered by spectral quality issues, resulting from intrinsic low \acp{snr}, artifacts, unknown baseline contributions, and other background signals. \cite{Hurd2009ArtifactsAPI} Furthermore, overlapping metabolites and limited spectral resolution, even at high field strengths, can affect the reliability of metabolite quantification. \cite{kreis_terminology_2021} Consequently, human experts are often required, acquisition times are prolonged, and experiments need to be repeated. \cite{Kreis2004IssuesOS} These challenges have led to the development of various methods to enhance spectral quality, ranging from numerous model-based approaches to more recently introduced data-driven models \cite{Sande2023AReviewOfML}.

Among these methods are traditional denoising techniques such as \ac{hsvd} \cite{Rowland2021ACO} and time-frequency transform-based techniques \cite{Doyle1994SIFTAP, Cancino2002SignalDN, Ahmed2005NewDS}, which focus on transforming the data to isolate and remove noise, or sliding window methods \cite{Rowland2021ACO} that smooth the data by averaging neighboring points. These approaches have the drawback of over-smoothing the signals, thereby removing relevant information and obscuring quantification accuracy. Another conventional approach involves using total variation regularization for \ac{mrs} \cite{Joshi2015DenoisingMR} to reduce noise while maintaining sharp signal transitions. \cite{Strong2003EdgepreservingAS} This technique defines denoising as an optimization problem with the goal of finding a signal approximation that minimizes the total variation, e.g. the sum of squared errors. \cite{Joshi2015DenoisingMR} However, computational complexity is high and the approach's effectiveness depends on the optimal selection of the regularization parameter. \cite{Osadebey2014OptimalSO}

% Another approach called total variation defines denoising as an optimization problem trying to find a signal approximation that minimizes the total variation, e.g. the sum of squared errors. \cite{Joshi2015DenoisingMR} In this context, the goal is to maintain a high \ac{snr} while simultaneously ensuring the integrity of the signal and minimizing the occurrence of ringing artifacts. 

Alternatively, several deep learning techniques have been developed for data-driven denoising in \ac{mrs}, particularly using U-Net architectures \cite{lee_reconstruction_2020, Dziadosz2023DenoisingSM, Berto2024ResultsOT2023ISBI}, autoencoders \cite{lei_deep_2021, Wang2023DenoisingMR}, \acl{lstm} networks \cite{Chen2021MagneticRS}, and vision transformers \cite{Berto2024ResultsOT2023ISBI} to predict high \ac{snr} data from low-quality measurements. Although deep learning-based denoising has shown promising results in various applications, its data-dependency and black-box nature make it difficult to interpret the outcome and rely on the denoising process.

In addition to denoising, the detection and removal of artifacts is crucial to accurately estimate metabolite concentrations \cite{Tk2020WaterAL} and improve the reliability and repeatability of experiments \cite{Hurd2009ArtifactsAPI}. There has been recent progress in data-driven quality filtering \cite{pedrosa_de_barros_improving_2017, kyathanahally_quality_2018} and artifact detection \cite{gurbani_convolutional_2018, kyathanahally_deep_2018, jang_unsupervised_2021, hernandez-villegas_extraction_2022}, however, the removal of artifacts remains more challenging. An autoencoder has been proposed \cite{kyathanahally_deep_2018} to remove ghosting artifacts, also called spurious echoes, operating similarly to the denoising techniques \cite{lei_deep_2021, Wang2023DenoisingMR} by predicting clean data from corrupted data. An alternative approach \cite{lee_intact_2019} was introduced to remove any artifact by employing a \ac{cnn} architecture, taking contaminated spectra as input and predicting noise-free, metabolite-only spectra. However, these methods work directly and without oversight on the spectra, potentially altering them significantly. This can restrict the available fitting parameter space leading to systemic quantification offsets or even directly introduce metabolite concentration biases.

Wavelet analysis is commonly used as a signal quality enhancement technique and has shown promise in characterizing \ac{mrs} signals, capable of disentangling overlapping signal components and thereby removing noise, artifacts, and baseline contributions. \cite{Serrai1997TimeDomainQO, Suvichakorn2008AnalyzingMR} The basic principle of the wavelet transform is to represent a signal as a set of basis functions called wavelets. Common denoising techniques include wavelet thresholding and wavelet shrinkage. Wavelet thresholding \cite{Abramovich1998WaveletTV, Chang2000AdaptiveWT} involves zeroing out coefficients below a specific threshold, thereby eliminating noise or other unwanted components. In contrast, wavelet shrinkage \cite{Donoho1994IdealSA, Donoho1995AdaptingTU} reduces the magnitude of the wavelet coefficients, leading to more robust data representation and noise reduction. In the context of \ac{mrs}, wavelet transform has been used for various purposes \cite{Suvichakorn2010WaveletbasedTI}, such as quantification of signal parameters \cite{Serrai1997TimeDomainQO}, or characterization of non-parameterizable signal components \cite{Young1998AutomatedSA}, including baseline \cite{Soher2001RepresentationOS} and macromolecular contributions \cite{Suvichakorn2008AnalyzingMR}. A recent data-driven method \cite{Ji2021SpectralWA} employed a \acl{svm} to classify coefficients to selectively remove noise components while preserving metabolite peaks. This exploits the enhanced separability of \ac{mrs} signal components in the wavelet domain through a sample-adaptive approach. Nevertheless, metabolite signals exhibit high-frequency components that overlap with each other and noise, even within the wavelet domain. This is particularly pronounced at points of abrupt signal changes, necessitating a more flexible technique.

This work introduces the \ac{wand}, a novel physics-driven method for decomposing signals into desired individual components. By combining the power of wavelet analysis and the adaptability of neural networks, we achieve a comprehensive breakdown of \ac{mrs} signals into their metabolite-specific, baseline, and artifact components. This enables the targeted removal of unwanted signal components, such as artifacts and baseline contributions, while preserving the integrity of the desired metabolite signals, ultimately leading to more accurate and reliable metabolite quantification.
The proposed \ac{wand} architecture features a U-Net that predicts soft masks for the wavelet coefficients derived from the \ac{cwt} of the signal. These masks are used to isolate the desired signal components in the wavelet domain, which are then inverse-transformed to obtain the separated signals. The network is optimized with simulated \ac{mrs} spectra using the \ac{mse} of predicted and ground truth decompositions. Additionally, an artifact mask for unknown and non-parametrizable signals is created by inverting the sum of all known signals and ensuring complete coverage of the wavelet coefficients with a softmax function. This process yields a mask for unpredictable signals which can be used to remove unknown and random artifacts from \ac{mrs} signals. 
Numerical results with simulated spectra demonstrate an accurate decomposition, with \ac{mse} values of predicted and ground truth decompositions ranging from 2.32e-4\ (± 1.2e-5) to 1.74e-6\ (± 4.9e-8) for the normalized real components. The artifact removal capabilities significantly enhance the quantification accuracy of linear combination fitting. We further validate the \ac{wand}’s effectiveness using data from the 2016 MRS Fitting Challenge \cite{marjanska_mrs_2021, Marjanska2022resultsFittingCha} and in-vivo experiments.

The rest of the article is organized as follows: Section~\ref{sec:methods} describes the assumed signal model and introduces the wavelet transform; Section~\ref{sec:models} introduces the proposed \ac{wand}, including details on neural architecture and training; Section~\ref{sec:material} introduces the simulated and in-vivo data; Section~\ref{sec:results} presents the results; Section~\ref{sec:discussion} and Section~\ref{sec:conclusion} provide a discussion and concluding remarks.

%%%%%%%%%%%%%%%
%%% Methods %%%
%%%%%%%%%%%%%%%
\section{System Model \& Preliminaries}\label{sec:methods}
% overview of section
% intro MRS signals, simulations, training data, etc.
% intro CWT

In this section, we detail the system model for which we train the \ac{wand} and introduce other prerequisites. With that aim, we first present the assumed \ac{mrs} signal model as well as the noise and artifact generation in Section~\ref{ssec:signal_model}. Then, we formulate the \ac{cwt} and discuss its application to \ac{mrs} in Section~\ref{ssec:cwt}.

\subsection{Signal Model}\label{ssec:signal_model}
% intro signal model/components

The chosen \ac{mrs} signal model follows the form of a standard Voigt-lineshaped model \cite{clarke_fslmrs_2021}: 
\begin{equation} \label{eq:metab_model}
    S_{m}(f) = e^{i (\phi_0 + f \phi_1)} a_m \mathcal{F}\{s_{m}(t) \ e^{- (\gamma + \varsigma^2 t + i\epsilon) t}\},
\end{equation}
\begin{equation} \label{eq:signal_model}
    X(f) = \sum^{M}_{m=1} S_{m}(f) + B(f) + R(f) + P(f) + N,
\end{equation}
where $\phi_0$ and $\phi_1$ are the zero and first-order phases, the $a_{m}$ are the concentrations of the $M$ metabolites of interest, $\gamma$ and $\varsigma$ are Lorentzian and Gaussian broadening, respectively, and $\epsilon$ represents a frequency shift. $\mathcal{F}\{\cdot\}$ denotes the Fourier transform and the basis functions $\{s_m(t)\}_{m=1}^M$ are the full spectral contributions of the individual metabolites, obtained using density matrix simulations \cite{simpson_advanced_2017, Zhang2017FastSpectroscopy}. \Ac{mm} are included, experiencing the same broadening, phasing, and shifting as the metabolites. Furthermore, the baseline $B(f)$ is represented by a complex second-order polynomial, while $R(f)$, $P(f)$, and $N$ correspond to spectral distortions in form of random baselines, peaks, and \acl{awgn}, respectively. The random walk \cite{Rayleigh1905ThePO} $R(f)$ starts with a random initialization at $R(0)$ and takes steps of a given size in a random direction followed by a smoothing of the generated line that is bounded by a defined minimum and maximum. It is drawn separately for real and imaginary parts. The Gaussian peaks $P(f)$ are obtained as follows:
\begin{equation}
    P(f) = \sum^{M + K}_{k=M + 1} \mathcal{H}\{a_k e^{-\frac{(f - \epsilon_k)^2}{2 \varsigma_k^2}}\} e^{-i\phi_{0, k}},
\end{equation}
with $K$ being the number of peaks and $\mathcal{H}\{\cdot\}$ representing the Hilbert transform.

\subsection{Continuous Wavelet Transform}\label{ssec:cwt}
% intro wavelets
% brief overview of applications
% math...
% motivation & explanation for use in mrs

The principle of wavelet analysis \cite{Meyer1989WaveletsAO, Merry2005WaveletTA, Mallat2008AWT, Addison2016IllustratedWT} is to represent a signal as a superposition of wavelets which are generated from a single mother wavelet through the process of scaling and translating the signal. \cite{Graps1995AnIntroToWavelets} %There exist many applications based on the concept of wavelet analysis \cite{Akansu2010FullLA}, some of which are mentioned in Section~\ref{sec:intro}. A more extensive overview of applications and methodologies can be found in the following textbooks \cite{Meyer1989WaveletsAO, Merry2005WaveletTA, Mallat2008AWT, Addison2016IllustratedWT}.
The forward \ac{cwt} is defined as:
\begin{equation}\label{eq:cwt}
    W(s, \delta) = \frac{1}{\sqrt{s}} \int X(f) \Psi^*\left(\frac{f-\delta}{s}\right) \,df,
\end{equation}
where $s$ corresponds to the scale factor, $\delta$ is the translation parameter, and $\Psi(f)$ represents the mother wavelet (with $^*$ denoting its complex conjugate). \cite{Aguiar2011TCWT} This transformation yields a two-dimensional representation of the signal called scalogram $|W(s, \delta)|$. \cite{Torrence1998APG} The \ac{icwt} allows reconstruction of the signal and is obtained by integrating over all scales and translations of the wavelet coefficients:
\begin{equation}\label{eq:icwt}
    X(f) = C_\Psi^{-1} \iint W(s, \delta) \frac{1}{\sqrt{s}} \Psi\left(\frac{f-\delta}{s}\right) \, d\delta \, ds,
\end{equation}
with the admissibility constant $C_\Psi$ depending only on the wavelet $\Psi$. \cite{Daubechies1992TenLO} Due to the high redundancy of the transform many reconstruction formulas exist, and with the assumption of $\Psi$ being analytic and $X(f)$ real, we can simplify~\eqref{eq:icwt} to
\begin{equation}
    X^\Re(f) = \Re\left\{ C_\Psi^{-1} \int W(s, \delta) s^{-3/2} \,ds \right\},
\end{equation}
with $\Re\{\cdot\}$ representing the real part. \cite{Farge1992WaveletTA, Daubechies2011SynchrosqueezedWT}

\begin{figure*}
    \centering
    \hspace{-4mm}
    \includegraphics[width=1.85\columnwidth]{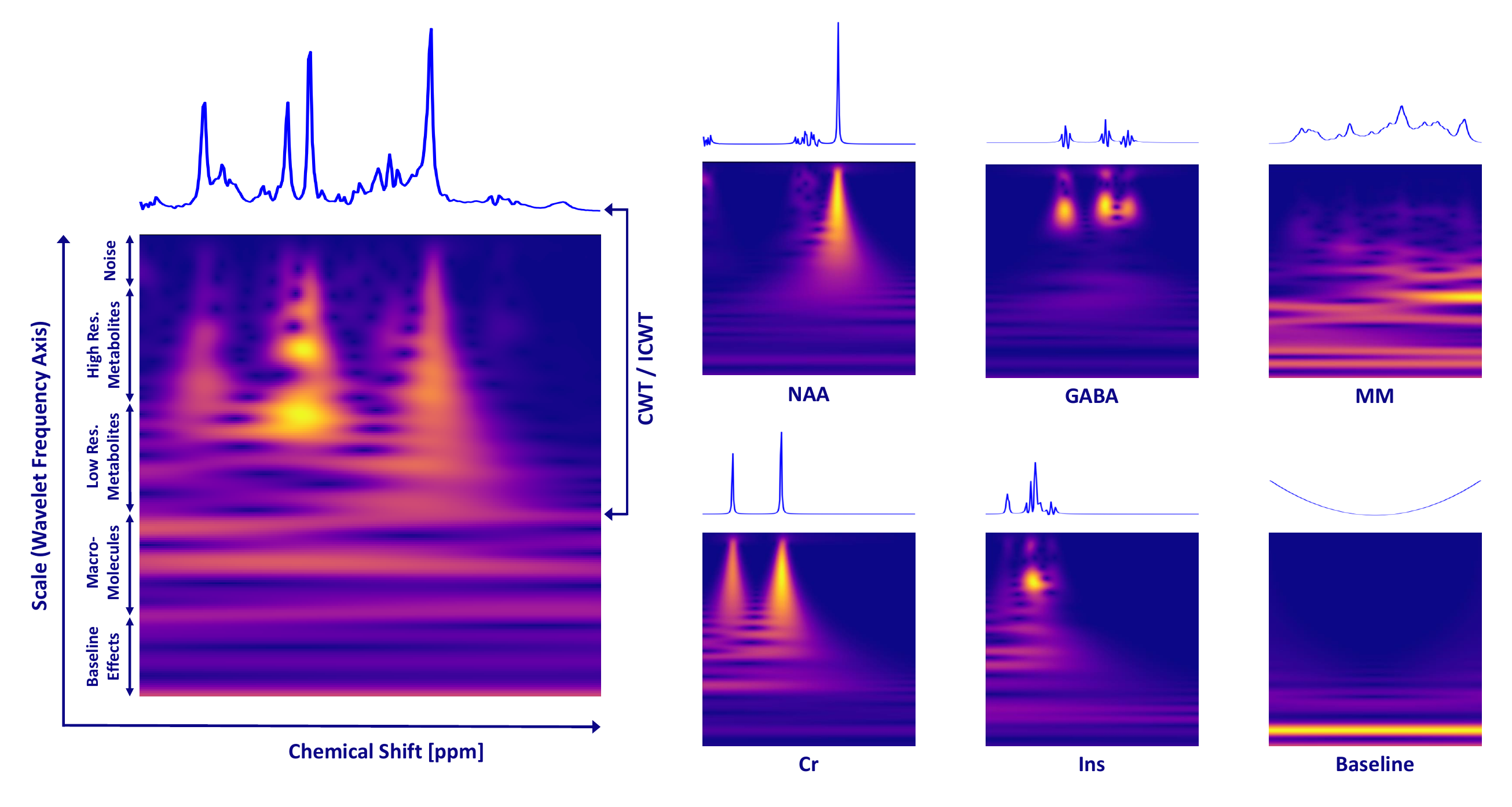}
    \hspace{4mm}
    \caption{Depicts the real part of a simulated example \ac{mrs} spectrum and its associated \ac{cwt}, along with spectra and scalograms of various metabolites and other signal components (\acs{naa}, \acs{gaba}, \acs{mm}, \acs{cr}, \acs{ins}, and a baseline). The scalograms demonstrate the enhanced separation of the metabolite signals in the wavelet domain, effectively allowing separation between high-frequency noise, metabolites, \acs{mm}, and other baseline components.} 
    \label{fig:intro_wavelets}
\end{figure*}

Figure~\ref{fig:intro_wavelets} shows the real part of an example spectrum $X^\Re(f)$ alongside its corresponding scalogram $|W^\Re(s, \delta)|$ as well as a selection of basis spectra $S_m^\Re(f)$ with their respective scalograms. %We utilize $^\Re$ and $^\Im$ to denote real and imaginary parts.  
Throughout this work, we employ the popular Morlet wavelet \cite{Aguiar2011TCWT} and apply the wavelet transform in the frequency domain to retain the well-known chemical shift axis for enhanced interpretability of the decomposition. Additionally, this allows truncation of the signal without losing any metabolite contributions by selecting only a relevant ppm range. Also, most \ac{mrs} quantification methods utilize the frequency domain for \ac{lcm} \cite{provencher_estimation_1993, Oeltzschner2020OspreyOP, clarke_fslmrs_2021}. However, it is not a necessity and the method also operates identically in the time domain.
%The Morlet wavelet, the most popular wavelet \cite{Aguiar2011TCWT}, is utilized for the transform and is constantly employed throughout this work. We perform the wavelet transform in the frequency domain, to retain the chemical shift axis while adding the additional scale axis (for more detailed discussion on the choice of the frequency domain see Section~\ref{ssec:discussion}). We observe enhanced separability in the wavelet domain of the overlapping signals, effectively differentiating between high-frequency noise, metabolites, macromolecular contributions, and other baseline effects.

%%%%%%%%%%%%%%
%%% Models %%%
%%%%%%%%%%%%%%
\section{Wavelet Analysis-based Neural Decomposition}\label{sec:models}
% intro to methodology
% detailed walk-through architecture
% discussion of the training procedure
% critical view on benefits, drawbacks, etc.

The following sections introduce the proposed \ac{wand}. Our approach achieves a data-driven decomposition of \ac{mrs} signals into metabolite-specific, baseline, and artifact components by taking advantage of their enhanced separability in the wavelet domain. Particularly, we train a neural network to mask the wavelet coefficients to obtain the desired signal components in the wavelet domain and then subsequently inverse transform them to attain the separated signals. We next elaborate on the architecture of the \ac{wand} in Section~\ref{ssec:arch} and then present the training method in Section~\ref{ssec:training}.

\clearpage
\begin{figure*}
    \centering
    % \hspace{-2mm}
    \includegraphics[width=2\columnwidth, trim= 0 0mm 0 0mm, clip]{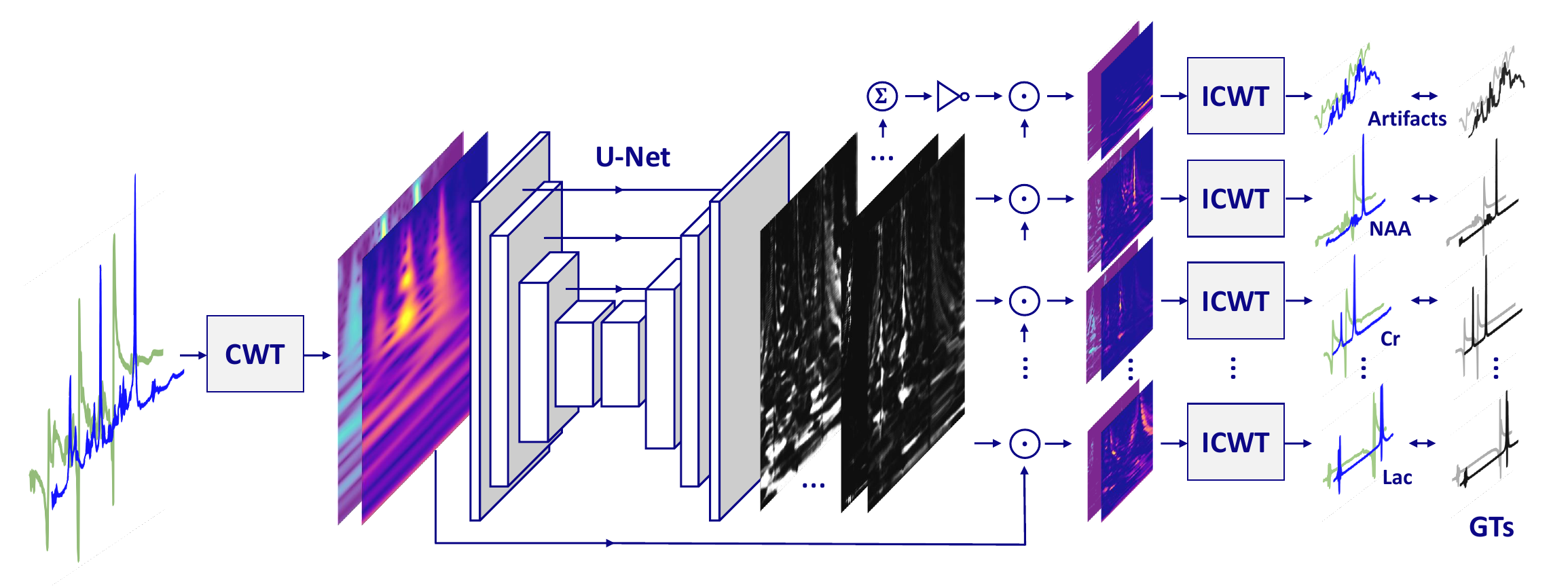}
    % \hspace{2mm}
    \caption{Schematic of the \ac{wand} architecture. The proposed network produces soft masks for the signal's real and imaginary components. These masks are applied to the wavelet coefficients, effectively isolating information specific to each metabolite. The resulting metabolite-specific scalograms can then be converted back into spectra using the \ac{icwt}. An artifact mask is created by summing all the predicted masks and subsequently inverting them. A softmax function is applied to ensure that the sum of all resulting masks equals one, thereby achieving an absolute decomposition of the signal, which is compared to a ground truth decomposition.
    }
    \label{fig:architecture}
\end{figure*}
\clearpage

\subsection{Architecture}\label{ssec:arch}
An overview of the \ac{wand} architecture is detailed in Figure~\ref{fig:architecture}. The pipeline starts by individually transforming the relevant frequencies of the real and imaginary parts of the spectrum into their respective two-dimensional \ac{cwt} representations. %Also, we remove the DC component from the input signal to reduce downstream reconstruction imperfections with the \ac{icwt} (see Section~\ref{ssec:discussion} for more details).
However, having a short signal with frequency extrema amplifies imperfections in the \ac{cwt} filterbank resulting in an imperfect reconstruction with the \ac{icwt}. \cite{Daubechies2011SynchrosqueezedWT} To alleviate the issue of an underrepresented zeroth filterbank, we remove the DC component from the signal prior to computing the \ac{cwt}. The effects of this on quantification are minimal and mostly depend on the initialization of the baseline modeling parameters of \ac{lcm} methods.
The stacked scalograms serve as an input to a U-Net, a \ac{cnn} architecture that is particularly effective for image segmentation \cite{Ronneberger2015UNetCN}, composed of an encoder and a decoder path with skip connections. See Appendix~\ref{app:details}, Table~\ref{tab:unet} for the specific neural network layer details. It outputs soft segmentation masks $\{H_m^\Re(s, \delta), H_m^\Im(s, \delta)\}_{m=1}^{M+1} \in [0, 1]$ for both the real and imaginary parts of all considered metabolites and a baseline component. Furthermore, an artifact mask is obtained by summing and inverting all these masks for real and imaginary parts, i.e. $H_{M+2}^\Re = 1 - \sum_{m=1}^{M+1} H_{m}^\Re$ and $H_{M+2}^\Im = 1 - \sum_{m=1}^{M+1} H_{m}^\Im$. After which all masks are pushed through a softmax activation to ensure they sum to one, resulting in the complete coverage of the wavelet domain representation and an absolute decomposition of the signal in both the real and imaginary parts. Finally, the masks are applied to the wavelet coefficients creating component-specific wavelet domain representations that can be converted back to individual spectra using the \ac{icwt}:
\begin{equation}
\begin{split}
    \hat{S}_m^\Re(f) &= \text{ICWT}\{W^\Re(s, \delta) \cdot H_m^\Re(s, \delta)\},\\
    \hat{S}_m^\Im(f) &= \text{ICWT}\{W^\Im(s, \delta) \cdot H_m^\Im(s, \delta)\}.
\end{split}
\end{equation}

\subsection{Training Strategy}\label{ssec:training}

\begin{table*}[b]%
\caption{Overview of the notations and distribution ranges of the simulation parameters.  The metabolite concentration ranges are taken from De Graaf 2019 \cite{de_graaf_vivo_2019}. $\mathcal{U}[a, b]$ and $\mathcal{U}\{a, b\}$ denote continuous and discrete uniform distributions, respectively. \label{tab:sim_params}}%
\vspace{2mm}
\begin{tabular*}{\columnwidth}{@{\extracolsep\fill\hspace{1mm}}lcc@{\extracolsep\fill\hspace{1mm}}}
\toprule
\textbf{Parameter}                     & \textbf{Notation} & \textbf{Range}                   \\
\midrule
Acetate (Ace)                          & \( a_1 \)         & \( \mathcal{U}[0.0, 0.5] \) \\
Alanine (Ala)                          & \( a_2 \)         & \( \mathcal{U}[0.1, 1.6] \)   \\
Ascorbate (Asc)                        & \( a_3 \)         & \( \mathcal{U}[0.4, 1.7] \)   \\
Aspartate (Asp)                        & \( a_4 \)         & \( \mathcal{U}[1.0, 2.0] \)   \\
Creatine (Cr)                          & \( a_5 \)         & \( \mathcal{U}[4.5, 10.5] \)   \\
Gamma-Aminobutyric Acid (GABA)         & \( a_6 \)         & \( \mathcal{U}[1.0, 2.0] \)   \\
Glycerophosphocholine (GPC)            & \( a_7 \)         & \( \mathcal{U}[0.4, 1.7] \)   \\
Glutathione (GSH)                      & \( a_8 \)         & \( \mathcal{U}[1.7, 3.0] \)   \\
Glucose (Glc)                          & \( a_9 \)         & \( \mathcal{U}[1.0, 2.0] \)   \\
Glutamine (Gln)                        & \( a_{10} \)      & \( \mathcal{U}[3.0, 6.0] \)  \\
Glutamate (Glu)                        & \( a_{11} \)      & \( \mathcal{U}[6.0, 12.5] \)  \\
Glycine (Gly)                          & \( a_{12} \)      & \( \mathcal{U}[0.2, 1.0] \)  \\
Myo-Inositol (Ins)                     & \( a_{13} \)      & \( \mathcal{U}[4.0, 9.0] \)  \\
Lactate (Lac)                          & \( a_{14} \)      & \( \mathcal{U}[0.2, 1.0] \)  \\
Macromolecules (MM)                    & \( a_{15} \)      & \( \mathcal{U}[0.0, 10.0] \)  \\
N-Acetylaspartate (NAA)                & \( a_{16} \)      & \( \mathcal{U}[7.5, 12.0] \)  \\
N-Acetylaspartylglutamate (NAAG)       & \( a_{17} \)      & \( \mathcal{U}[0.5, 2.5] \)   \\
Phosphocholine (PCho)                  & \( a_{18} \)      & \( \mathcal{U}[0.2, 1.0] \)   \\
Phosphocreatine (PCr)                  & \( a_{19} \)      & \( \mathcal{U}[3.0, 5.5] \)   \\
Phosphoethanolamine (PE)               & \( a_{20} \)      & \( \mathcal{U}[1.0, 2.0] \)   \\
Taurine (Tau)                          & \( a_{21} \)      & \( \mathcal{U}[3.0, 6.0] \)   \\
Scyllo-Inositol (sIns)                 & \( a_{22} \)      & \( \mathcal{U}[0.2, 0.5] \)   \\
\bottomrule
\end{tabular*}
\begin{tabular*}{\columnwidth}{@{\extracolsep\fill\hspace{1mm}}lcc@{\extracolsep\fill\hspace{1mm}}}
\toprule
\textbf{Parameter}                     & \textbf{Notation} & \textbf{Range}                   \\
\midrule
Number of Metabolites (+MM)            & \( M \)           & \( \{ 22 \} \) \\
Frequency Shifts                       & \( \epsilon \)    & \( \mathcal{U}[-5, 5] \) \\
Lorentzian Broadening                  & \( \gamma \)      & \( \mathcal{U}[2, 15] \)  \\
Gaussian Broadening                    & \( \varsigma \)   & \( \mathcal{U}[2, 15] \)  \\
Zeroth-Order Phase                     & \( \phi_0 \)      & \( \mathcal{U}[-0.2, 0.2] \) \\
First-Order Phase                      & \( \phi_1 \)      & \( \mathcal{U}[-10^{-5}, 10^{-5}] \) \\
Baseline Coefficients                  & \( b_1 \)         & \( \mathcal{U}[-60, 20] \)   \\
                                       & \( b_2 \)         & \( \mathcal{U}[-80, 40] \)   \\
                                       & \( b_3 \)         & \( \mathcal{U}[-100, 60] \)   \\
                                       & \( b_4 \)         & \( \mathcal{U}[-60, 100] \)   \\
                                       & \( b_5 \)         & \( \mathcal{U}[-160, 20] \)   \\
                                       & \( b_6 \)         & \( \mathcal{U}[-40, 100] \)   \\
Random Walk Step Size                  &                   & \( \mathcal{U}[0, 10^3] \)   \\
Random Walk Smoothing                  &                   & \( \mathcal{U}[1, 10^3] \)   \\
Random Walk Min. Bound                 &                   & \( \mathcal{U}[-10^4, 0] \)   \\
Random Walk Max. Bound                 &                   & \( \mathcal{U}[0, 10^4] \)   \\
Number of Peaks                        & \( K \)           & \( \mathcal{U}\{0, 5\} \)   \\
Peak Amplitudes                        & \( a_{M+1}, ..., a_{M+K} \)         & \( \mathcal{U}[0, 6 \cdot 10^4] \)   \\
Peak Widths                            & \( \varsigma_{M+1}, ..., \varsigma_{M+K} \) & \( \mathcal{U}[0, 100] \)   \\
Peak Phases                            & \( \phi_{M+1}, ..., \phi_{M+K} \)      & \( \mathcal{U}[0, 2 \pi] \)   \\
Peak Locations                         & \( \epsilon_{M+1}, ..., \epsilon_{M+K} \)  & \( \mathcal{U}\{0, 2048\} \)   \\
Complex Gaussian Noise$^{*}$           & \( N \)           & \( \mathcal{CN}(0, 500) \) \\
\bottomrule
\end{tabular*}
\begin{tablenotes}%%[341pt]
\item[$^{*}$] \Ac{snr} ranging from 5 - 25 dB as computed from the ground truth metabolite and noise signals over 0.5 to 4.5 ppm.
\end{tablenotes}
\end{table*}

The \ac{wand} is trained end-to-end in a supervised setting as a regression problem with simulated \ac{mrs} data. To increase data variability, we simulate each batch ad-hoc during training by drawing the signal model parameters from uniform distributions (see Table~\ref{tab:sim_params} for details) and employing equations~\eqref{eq:metab_model} and~\eqref{eq:signal_model} to obtain a batch of spectra $\{X_u, Y_u\}_{u=1}^U$ with $U$ samples and ground truths $Y_u = (S_1, ..., S_M, S_{M+1}=B, S_{M+2}=R + P + N)_u$, which hold the individual signal components before the summation in Equation~\eqref{eq:signal_model}. Note that we combine the contamination's $R, P, N$ into on single artifact component $S_{M+2}$.

Having the ground truths of the simulations, we can compare the predicted decomposition $\hat{Y}_u$ with actual decomposition $Y_u$ via the \ac{mse} of the individual spectra:
\begin{equation}
\begin{split}
    \mathcal{L}(Y_u, \hat{Y}_u) = \frac{1}{2(M+2)} (\sum_{m=1}^{M+2} \text{MSE}(S_m^\Re(f), \hat{S}_m^\Re(f)) \\
    + \sum_{m=1}^{M+2} \text{MSE}(S_m^\Im(f), \hat{S}_m^\Im(f))).
\end{split}
\end{equation}
The \ac{mse} is computed over a defined frequency range, $f_{min} \leq f \leq f_{max}$, corresponding to 0.5 to 4.5 ppm throughout this work.
After loss computation, we perform batch gradient descent using the Adam optimizer to update the weights of the U-Net via backpropagation through the architecture detailed in Figure~\ref{fig:architecture}. See Appendix~\ref{app:details}, Table~\ref{tab:ml_params} for a detailed overview of the relevant training parameters.

%%%%%%%%%%%%%
%%% Setup %%%
%%%%%%%%%%%%%
\section{Materials \& Methods} \label{sec:material}
The following section introduces the data and analysis methods used to evaluate the performance of the proposed \ac{wand} method. We utilize two different sets of artificial data and two different types of in-vivo data, as well as two different analysis methods for metabolite quantification with and without the use of \ac{wand} for artifact removal.

\subsection{Data Setup} \label{ssec:data_setup}
The first type of simulated data is generated equally to the training data (described in Section \ref{ssec:training}) by drawing the signal parameters of equations \eqref{eq:metab_model} and \eqref{eq:signal_model} from uniform distributions as detailed in Table~\ref{tab:sim_params}. This allowed the creation of a wide range of spectra with varying levels of metabolite concentrations, line broadening, phase shifts, and simulated artifacts. 

The second set of artificial data consists of the 28 samples provided in the 2016 MRS Fitting Challenge data~ \cite{marjanska_mrs_2021, Marjanska2022resultsFittingCha}. This data includes spectra with varying \acp{snr}, lineshapes (Lorentzian and Voigt), linewidths, \acs{gaba} and \acs{gsh} concentrations, macromolecular contributions, artifacts such as eddy currents and residual water signals, as well as samples mimicking tumor spectra. \cite{Marjanska2022resultsFittingCha}
Both types were generated using the same basis set, based on the \ac{press} sequence with the following parameters: TE = 30~ms, TE1 = 11~ms, TE2 = 19~ms, TR >> T1, 123.22~MHz center frequency, 4000~Hz bandwidth, and 2048 spectral points.

The first in-vivo data samples are from the publicly available "fMRS in pain" data from Archibald et al.~\cite{Archibald2020MetaboliteAI}, acquired at 3T with \ac{press} and the following parameters: TE = 22~ms, TR = 4000~ms, 127.795~MHz center frequency, 2000~Hz bandwidth, and 2048 spectral points. We average the 15 baseline acquisitions for different \ac{nsa} to create spectra with varying levels of noise. 

The second type of in-vivo data consists of spectra acquired from a healthy volunteer, who gave written informed consent, at Philips Medical System International B.V., Best, The Netherlands. It consists of two different voxel locations near the skull, with one voxel specifically chosen to include lipid contamination from subcutaneous fat. The acquisition parameters are: 3~T, \ac{press}, TE = 30~ms, TR = 4000~ms, 127.75~MHz center frequency, 4000~Hz bandwidth, and 2048 spectral points. See Appendix~\ref{app:mrs_in_mrs}, Table \ref{tab:mrsinmrs_frms} and \ref{tab:mrsinmrs_lipids}, for the \ac{mrsinmrs} \cite{lin_minimum_2021}.

\subsection{Processing \& Analysis} \label{ssec:analysis}
% mention WAND being trained for all acquisition types by using the corresponding basis set for the simualtions
% simulated data as is (no processing needed, everything incorporated in the signal model)
% in-vivo data processed with FSL-MRS (see MRSinMRS for details)
% FSL-MRS and LCModel used for fitting
% introduce optimal referencing
The neural network of the proposed \ac{wand} method is trained separately for each acquisition type by adjusting the basis set for the training data accordingly. Simulated data is used directly without any additional processing, as all necessary components are incorporated within the signal model. In contrast, the in-vivo data is processed using FSL-MRS (see \ac{mrsinmrs} in Appendix \ref{app:details} for details). Metabolite quantification is performed using two \ac{lcm} fitting methods: FSL-MRS (Newton method) \cite{clarke_fslmrs_2021} and LCModel \cite{provencher_estimation_1993}. Fitting is conducted in the default range of LCModel, from 0.5 to 4.2 ppm, for all methods throughout this work. Furthermore, absolute quantification for the simulated data is obtained from relative metabolite concentration estimates $\{\hat{a}_m\}_{m=1}^M$ by optimally scaling to the absolute ground truth values $\{a_m\}_{m=1}^M$ using
\begin{equation} \label{eq:optimal_ref}
    w_{opt} = \min_w \sum_{m=1}^M \left |w \ \hat{a}_m - a_m\right |,
\end{equation}
leading to an error computation via $\left |w_{opt} \ \hat{a}_m - a_m\right |$ for all metabolites $M$.
This ensures an equal comparison of concentration estimates without a water reference or the influence of a single estimate, such as when referencing to e.g. \ac{tcr}.

%%%%%%%%%%%%%%%
%%% Results %%%
%%%%%%%%%%%%%%%
\section{Results}\label{sec:results}
In the following section, we present our experimental evaluations of the \ac{wand} method. First, we showcase the achievable decompositions as well as benchmark \ac{wand} against common denoising techniques on a synthetic test setup corresponding to the configurations of the training data ranges (Section~\ref{sssec:simulations}); then analyze the potential to enhance quantification accuracy on the artificial data of the 2016 MRS Fitting Challenge \cite{marjanska_mrs_2021, Marjanska2022resultsFittingCha} (Section~\ref{sssec:challenge}); and finally validate the robustness of \ac{wand} on in-vivo experiments (Section~\ref{sssec:noise} and Section~\ref{sssec:lipids}).

% \subsection{Artificial Data}\label{ssec:artifical_data}
%In both Section~\ref{sssec:simulations} and~\ref{sssec:challenge} human brain spectra at 3~T were simulated for the \ac{press} sequence (TE = 30~ms, TE1 = 11~ms, TE2 = 19~ms, TR >> T1, 123.22~MHz center frequency, 4000~Hz bandwidth, 2048 spectral points). The simulated data of Section~\ref{sssec:simulations} was obtained as described in Section~\ref{ssec:training}, while the data of Section~\ref{sssec:challenge} corresponds to the 28 samples of the 2016 MRS Fitting Challenge data \cite{marjanska_mrs_2021}.

\subsection{Simulated Data}\label{sssec:simulations}
\begin{figure*}
    \centering
    \includegraphics[width=1.85\columnwidth, trim=0 0mm 0 0mm, clip]{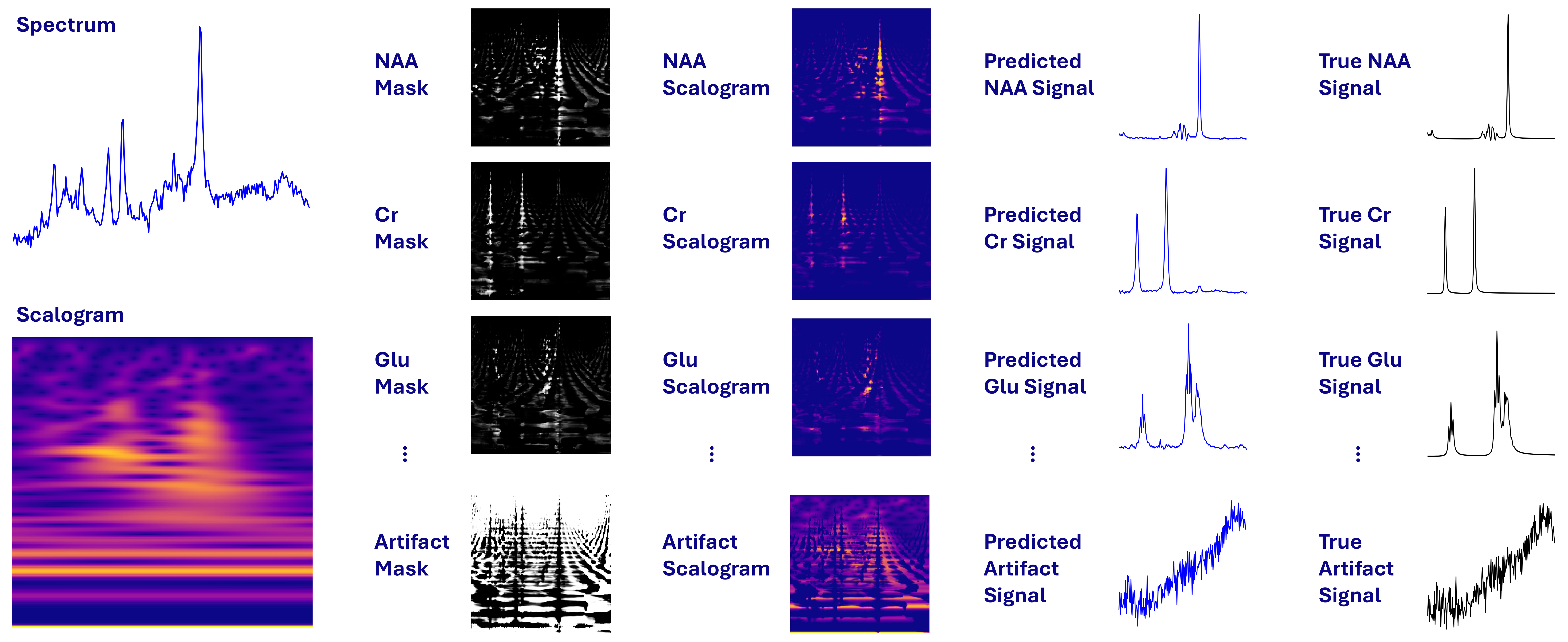}
    \caption{An illustrative summary of the steps involved within the \ac{wand} framework, showcasing the real part of a random spectrum, its associated scalogram, several predicted masks, their respective masked scalograms, and the ultimately obtained spectral components along the corresponding ground truth signals.} \label{fig:synth_overview}
\end{figure*}

We first provide an overview of the input, intermediate, and output steps involved in the \ac{wand} framework for the real component channels (the imaginary channels are placed in Appendix~\ref{app:add_materials}, Figure~\ref{fig:synth_overview_imag}). Figure~\ref{fig:synth_overview} shows a randomly selected test spectrum alongside its scalogram. This is followed by the masks created by the U-Net and the separate scalograms obtained after applications of the masks to the input scalogram. The spectra resulting from the \ac{icwt} of the decomposed scalograms are then displayed, along with the corresponding ground truth spectra of the individual metabolites for comparison. The predicted masks capture the fundamental signal attributes, including correct broadening, phase and frequency alignment.

\begin{figure*}
    \centering
    \includegraphics[width=2\columnwidth, trim=0 0mm 0 0mm, clip]{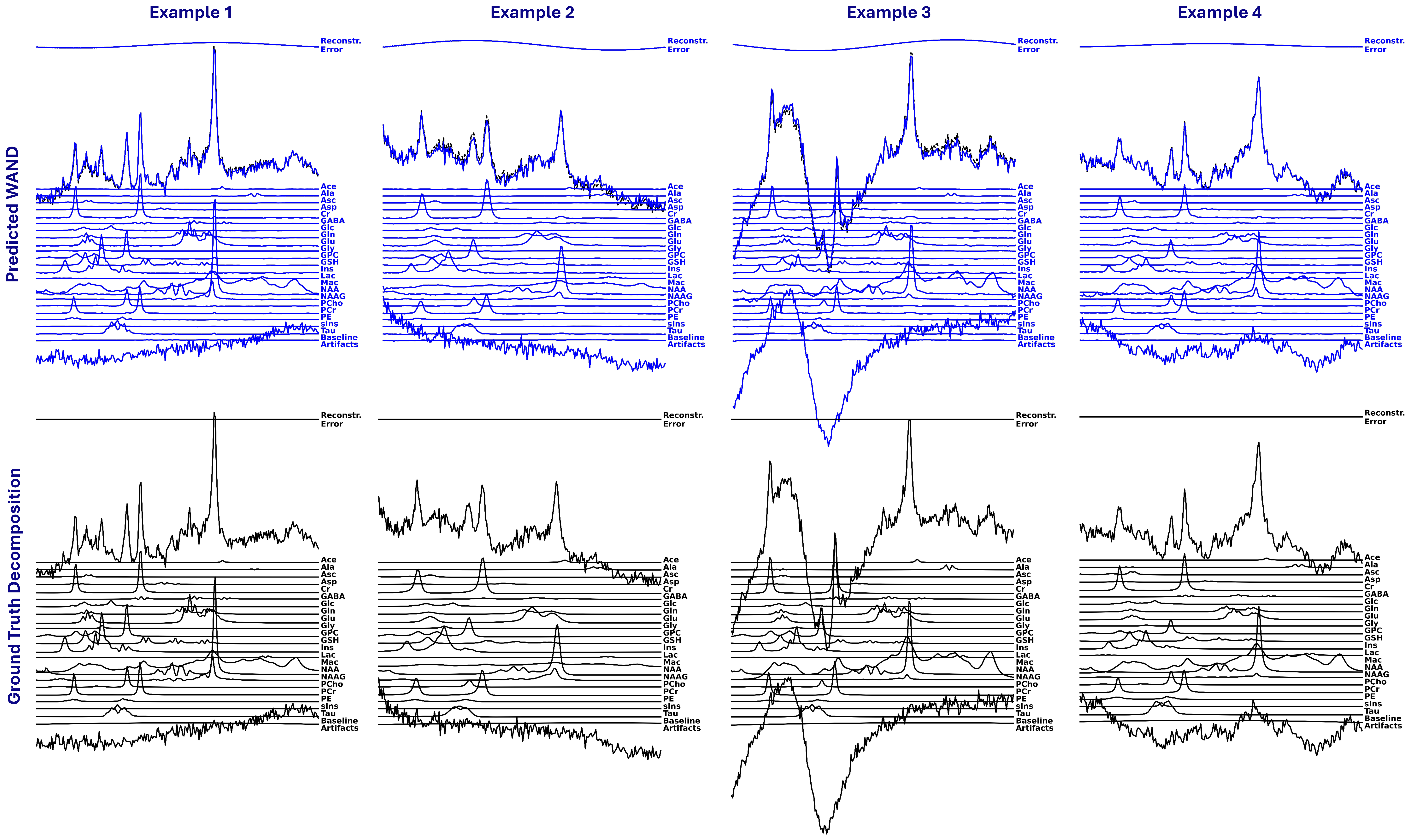}
    \caption{Shows the real parts of the complete predicted decomposition for four randomly selected examples, alongside their corresponding ground truth decompositions in the range of 0.5 to 4.5 ppm. The reconstruction error represents the residual of the original signal and the summed \ac{icwt} reconstructions.}
    \label{fig:synth_decomps}
\end{figure*}

Next, we present a more in-depth analysis of the complete decompositions achieved with \ac{wand}. Figure~\ref{fig:synth_decomps} depicts the real parts of the predicted \acp{wand} of four random example spectra alongside the corresponding ground truth decompositions in the chemical shift range of 0.5 to 4.5 ppm. The analogous decompositions of the imaginary parts are in Appendix~\ref{app:add_materials}, Figure~\ref{fig:synth_decomps_imag}. The reconstruction error represents the residual (of the real part) of the original signal and the summed \ac{icwt} reconstructions, i.e. $X - \sum_{m=1}^{M+2} S_m$, and arises from minor imperfections of the \ac{icwt}, as discussed in Section~\ref{ssec:arch}. The metabolite signals are effectively isolated by \ac{wand}, despite minor signal imperfections, resulting in Gaussian noise in the artifact channel that closely mirrors the actual ground truth noise signal.

To evaluate the accuracy of the predicted decomposition, we list the obtained \ac{mse} values when comparing the predicted and ground truth signal components of the decompositions in Table~\ref{tab:decomp}. The \ac{mse} is calculated separately for the real and imaginary parts of 1000 test sample spectra, which have amplitudes normalized to an absolute value of 1. The \ac{se} is indicated in brackets and alternative error metrics are provided in Appendix~\ref{app:add_materials}, Table~\ref{tab:decomp_all}.

\begin{table}[b]
    \centering
    \caption{\Ac{mse} (± \ac{se}) of the real and imaginary parts of the predicted and ground truth signal components, computed over a test set of 1000 random samples (with spectra normalized to an absolute value of 1). \label{tab:decomp}}%
    \vspace{2mm}
    \begin{tabular*}{\columnwidth}{@{\extracolsep{\fill}\hspace{1mm}}lllllllll@{\extracolsep{\fill}\hspace{1mm}}}
    \toprule
        \textbf{Component} & \textbf{MSE Real} & \textbf{MSE Imag} \\
    \midrule
        Ace & 1.75e-6\ (± 4.9e-8) & 2.41e-6\ (± 5.8e-8) \\ 
        Ala & 2.62e-6\ (± 6.3e-8) & 3.44e-6\ (± 7.4e-8) \\ 
        Asc & 2.53e-6\ (± 6.2e-8) & 3.30e-6\ (± 9.0e-8) \\ 
        Asp & 1.74e-6\ (± 4.9e-8) & 2.70e-6\ (± 5.7e-8) \\ 
        Cr & 2.61e-5\ (± 1.0e-6) & 3.06e-5\ (± 1.3e-6) \\ 
        GABA & 2.70e-6\ (± 5.7e-8) & 3.29e-6\ (± 7.0e-8) \\ 
        Glc & 3.52e-6\ (± 9.3e-8) & 5.42e-6\ (± 2.4e-7) \\ 
        Gln & 7.07e-6\ (± 1.9e-7) & 8.52e-6\ (± 3.1e-7) \\ 
        Glu & 1.51e-5\ (± 6.0e-7) & 1.80e-5\ (± 7.4e-7) \\ 
        Gly & 2.02e-6\ (± 6.4e-8) & 2.60e-6\ (± 6.3e-8) \\ 
        GPC & 1.46e-5\ (± 6.8e-7) & 1.54e-5\ (± 6.5e-7) \\ 
        GSH & 5.70e-6\ (± 1.5e-7) & 1.01e-5\ (± 5.1e-7) \\ 
        Ins & 1.54e-5\ (± 6.8e-7) & 1.56e-5\ (± 6.4e-7) \\ 
        Lac & 2.06e-6\ (± 5.4e-8) & 2.70e-6\ (± 5.8e-8) \\ 
        Mac & 3.72e-5\ (± 1.7e-6) & 6.51e-5\ (± 4.6e-6) \\ 
        NAA & 1.90e-5\ (± 6.2e-7) & 2.49e-5\ (± 9.0e-7) \\ 
        NAAG & 1.09e-5\ (± 4.1e-7) & 1.17e-5\ (± 4.5e-7) \\ 
        PCho & 1.09e-5\ (± 4.6e-7) & 1.27e-5\ (± 5.4e-7) \\ 
        PCr & 2.01e-5\ (± 8.3e-7) & 2.74e-5\ (± 1.2e-6) \\ 
        PE & 2.82e-6\ (± 7.3e-8) & 3.67e-6\ (± 1.3e-7) \\ 
        sIns & 2.16e-6\ (± 6.3e-8) & 3.15e-6\ (± 8.0e-8) \\ 
        Tau & 1.11e-5\ (± 5.2e-7) & 1.38e-5\ (± 6.0e-7) \\ 
        Baseline & 8.21e-6\ (± 2.4e-7) & 8.03e-6\ (± 2.4e-7) \\ 
        Artifacts & 2.32e-4\ (± 1.2e-5) & 4.85e-4\ (± 3.6e-5) \\ 
    \bottomrule
    \end{tabular*}
\end{table}

\begin{figure*}
    \centering
    \includegraphics[width=1.95\columnwidth, trim=0 2mm 0 2mm, clip]{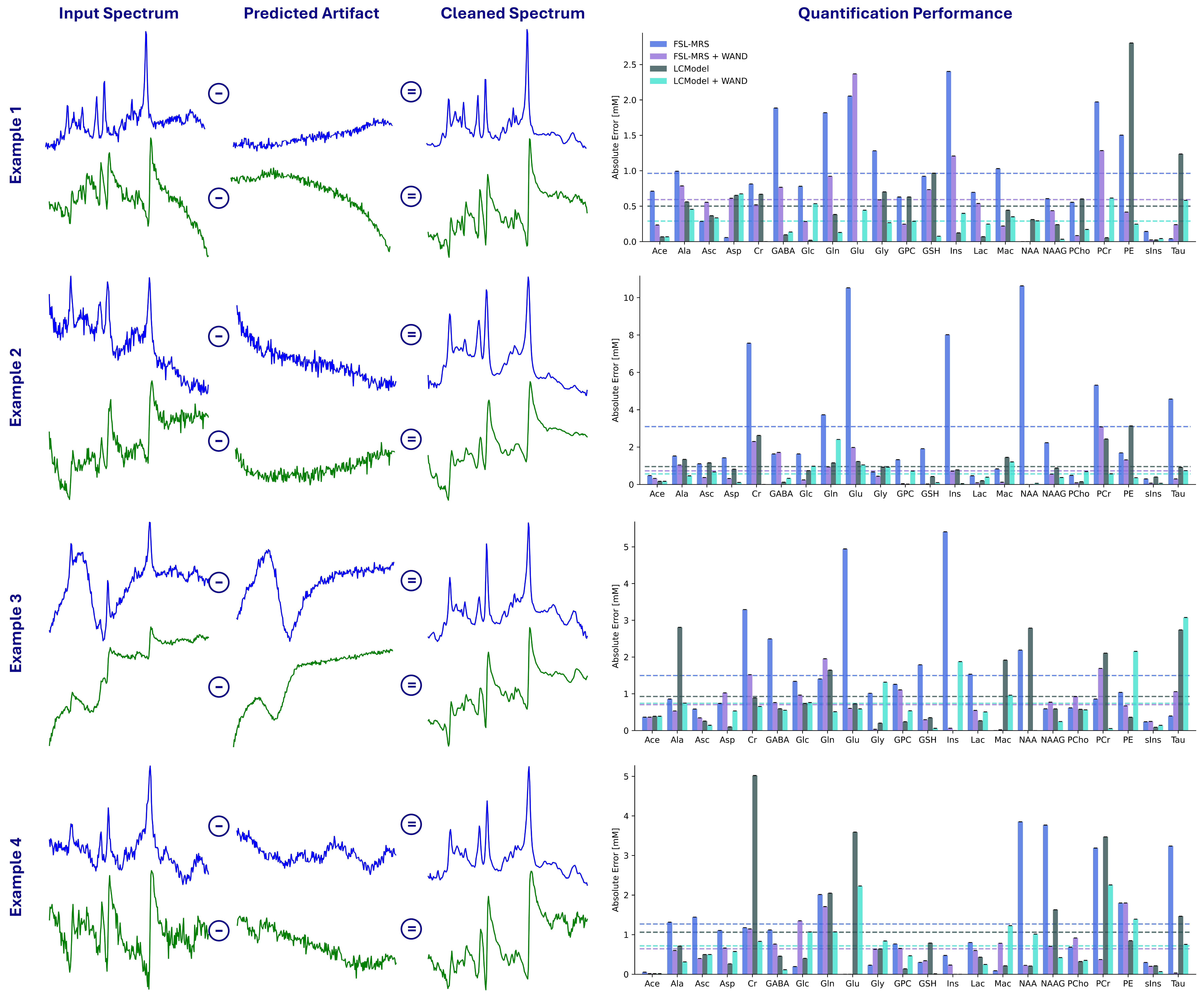}
    \caption{Depicts the real and imaginary parts of four example spectra, the predicted artifact component of the \ac{wand}, and the cleaned signal obtained by subtracting the artifact. Alongside the visualizations, the absolute quantification errors per metabolite are shown when comparing the input spectra to the cleaned spectra for both FSL-MRS and LCModel (\ac{mae} over all metabolites is provided through the dashed lines).}
    % \vspace{-3mm}
    \label{fig:synth_art_removal}
\end{figure*}

Given the isolation of metabolite and artifact signals through the decomposition, we can remove the artifact signal component from the input spectrum to obtain a cleaned, artifact-free spectrum. Figure~\ref{fig:synth_art_removal} depicts the achievable quality enhancement for the real and imaginary parts of the four random examples and additionally shows the quantification errors for LCModel and FSL-MRS with and without the use of \ac{wand} for artifact removal. We utilize Equation~\eqref{eq:optimal_ref} to obtain an optimal reference scaling from relative to absolute concentrations. 

\begin{figure*}
    \centering
    \includegraphics[width=2\columnwidth,  trim=0 2mm 0 2mm, clip]{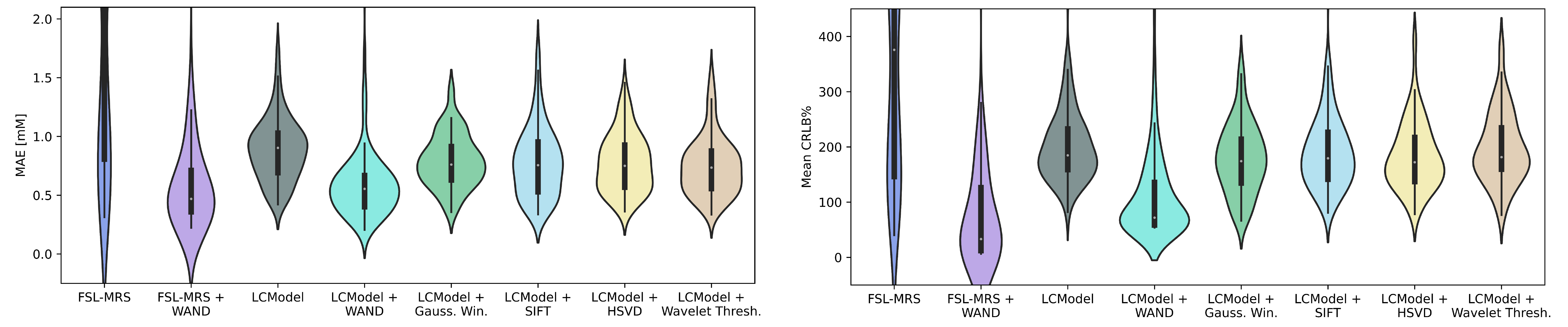}
    \caption{Comparison of the quantification \ac{mae} and mean \ac{crlbp} values over a test set of 100 samples for various methods: FLS-MRS standalone, FSL-MRS with \ac{wand}, LCModel standalone, LCModel with \ac{wand}, LCModel with a sliding Gaussian window, LCModel with \ac{sift}, LCModel with \ac{hsvd}, and LCModel with wavelet thresholding.}
    % \vspace{-3mm}
    \label{fig:synth_comparison}
\end{figure*}

In Figure~\ref{fig:synth_comparison}, we compare the performance of various processing and quantification methods over a test set of 100 sample spectra. Presented are the obtained \acp{mae} and corresponding \acp{crlbp} for FSL-MRS standalone, FSL-MRS in combination with \ac{wand} for removal of artifacts, LCModel by itself, LCModel with \ac{wand}, LCModel with a sliding Gaussian window for denoising, LCModel with \ac{sift} \cite{Doyle1994SIFTAP}, LCModel with \ac{hsvd} \cite{Rowland2021ACO}, and LCModel with wavelet thresholding \cite{Cancino2002SignalDN}. FLS-MRS quantification is strongly hindered by the artifact contributions in the signal, obtaining an average error of 1.92 (± 0.25), while in combination with \ac{wand} it achieves an average error of 0.66 (± 0.11). LCModel has larger baseline modeling freedom and achieves an average error of 0.89 (± 0.10) without preprocessing of the spectra and an error of 0.60 (± 0.07) with \ac{wand}. Furthermore, preprocessing the spectra with a Gaussian sliding window \cite{Rowland2021ACO}, \ac{sift} \cite{Doyle1994SIFTAP}, \ac{hsvd} \cite{Rowland2021ACO}, or wavelet thresholding \cite{Cancino2002SignalDN} leads to average LCModel quantification errors of 0.79 (± 0.08), 0.78 (± 0.09), 0.78 (± 0.09), 0.75 (± 0.08) respectively.

\clearpage
\clearpage

\clearpage
\clearpage

\subsection{2016 MRS Fitting Challenge}\label{sssec:challenge}

\begin{figure*}
    \centering
    \includegraphics[width=2\columnwidth,  trim=0 2mm 0 2mm, clip]{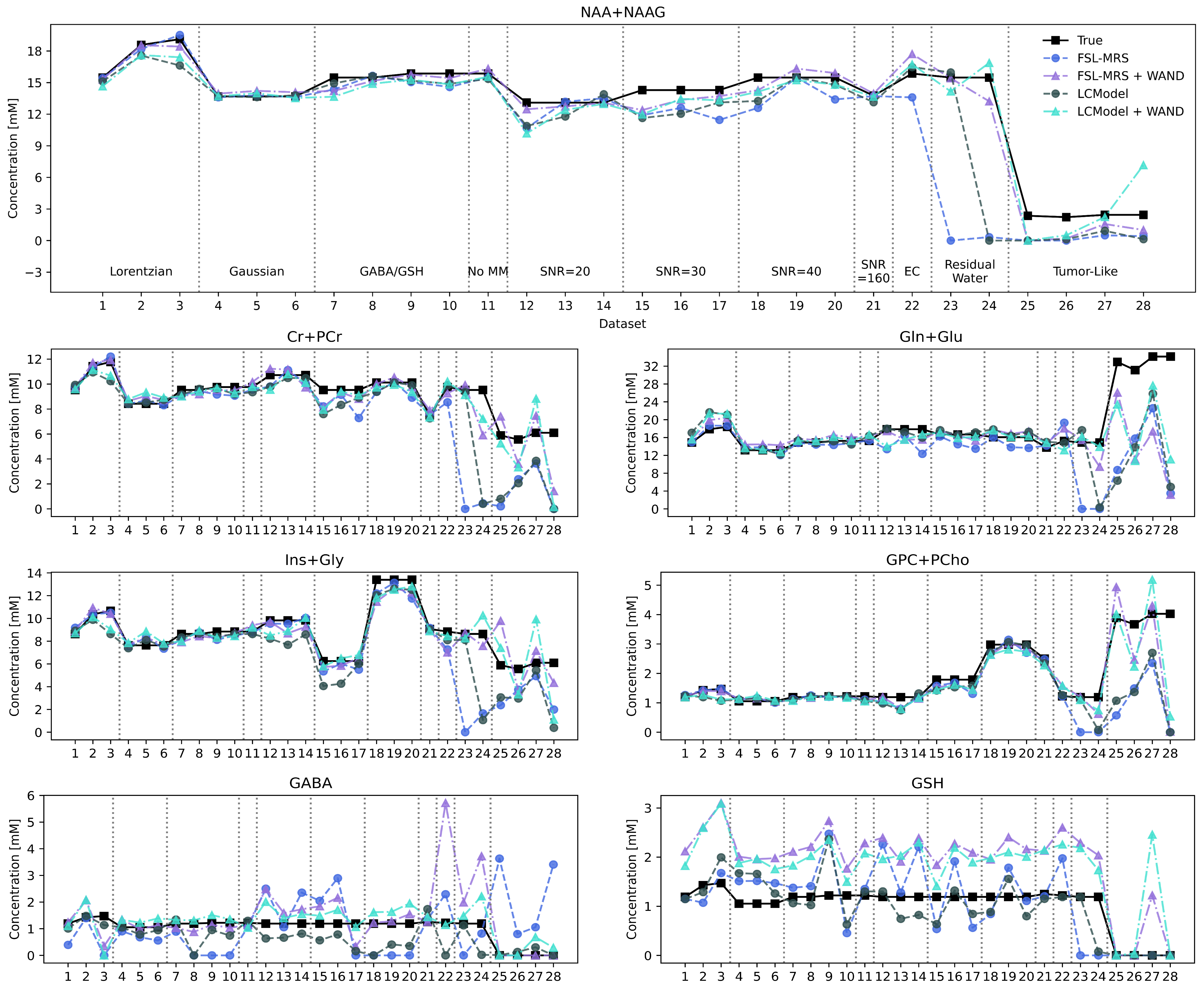}
    \caption{Comparison of standalone FSL-MRS and LCModel concentration estimates and in combination with \ac{wand} processed spectra for each of the 28 samples of the 2016 MRS Fitting Challenge data set for selected metabolites.
    %the major metabolites as well as \ac{gaba} and \ac{gsh}.
    }
    \label{fig:challenge_conc}
\end{figure*}

In this section, we validate the results achieved with \ac{wand} using the spectra of the 2016 MRS Fitting Challenge data set. Figure~\ref{fig:challenge_conc} provides an overview of the concentration estimates for selected metabolites, comparing the performance of standalone FSL-MRS and LCModel with their performance when combined with \ac{wand}. Again, we utilize Equation~\eqref{eq:optimal_ref} to obtain an optimal reference scaling from relative to absolute concentrations. The comparison highlights the impact of \ac{wand} on refining the concentration estimates, providing a clearer understanding of the metabolites’ presence in the samples, particularly for cases with strong artifacts such as residual water. For \acs{gsh}, we observed a systemic bias introduced by \ac{wand}, likely due to its training range of 1.7 to 3 mM (see Table~\ref{tab:sim_params}). However, it is noteworthy that a metabolite concentration of zero was correctly achieved in dataset samples 25, 26, and 28.

%The figure underscores the potential improvements in accuracy when \ac{wand} is utilized alongside traditional methods.

\begin{figure*}
    \centering
    \includegraphics[width=2\columnwidth,  trim=0 2mm 0 2mm, clip]{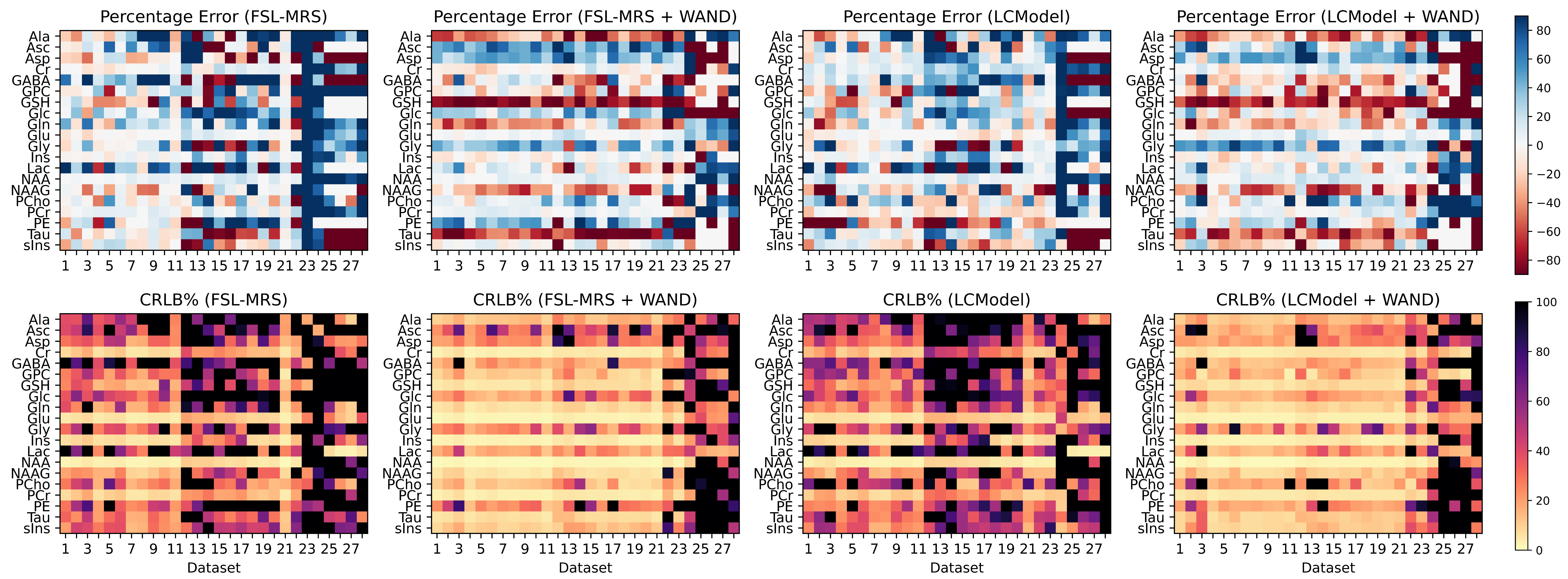}
    \caption{
    Depicts the obtained percentage error and \ac{crlbp} values for all 28 challenge samples for all concentration estimates of FLS-MRS and LCModel alone as well as in combination with \ac{wand}. %Red corresponds to an overestimation and blue for an underestimation of the predicted concentrations compared to the ground truth concentrations.
    }
    % \vspace{-3mm}
    \label{fig:challenge_perc}
\end{figure*}

Figure~\ref{fig:challenge_perc} provides a detailed overview, listing the percentage error and \acp{crlbp} values for all metabolites involved (excluding \acs{ace} and \acs{mm}) across the 28 challenge samples. Red corresponds to an overestimation and blue for an underestimation of the predicted concentrations compared to the ground truth concentrations. This figure highlights the concentration estimates obtained from FSL-MRS and LCModel, both independently and in conjunction with \ac{wand}. %The visual representation not only reveals observable biases but also shows the correlation with the \ac{crlbp} values. It offers valuabl insights into the accuracy and reliability of each method.
Although biases appear when using \ac{wand}, particularly for minor metabolites like \acs{ace}, \acs{gsh}, \acs{gly}, and \acs{pe}, we also observe a good correlation with the \ac{crlbp} values and generally enhanced precision for other metabolites.

% \subsection{In-Vivo Data}\label{ssec:in_vivo}
%In the following sections, we evaluate the performance of \ac{wand} with in-vivo data at 3 T using the \ac{press} sequence (see Appendix~\ref{app:mrs_in_mrs} for the \ac{mrsinmrs} \cite{lin_minimum_2021}). In Section~\ref{sssec:noise}, we utilize the publicly available data introduced by the work of Archibald et al. \cite{Archibald2020MetaboliteAI}, termed fMRS in pain data. While the data of Section~\ref{sssec:lipids} were acquired from healthy volunteers, who all gave written informed consent, at Philips Medical Systems International B.V., Best, The Netherlands. 

% fMRS in pain - \cite{Archibald2020MetaboliteAI}

% Metabolite quantification is in agreement with previously published values \cite{Pouwels1998RegionalMC, Baker2008RegionalAM, Minati2010QuantitationON}.

\subsection{Noise Interference} \label{sssec:noise}
\begin{figure*}
    \centering
    \includegraphics[width=1.95\columnwidth]{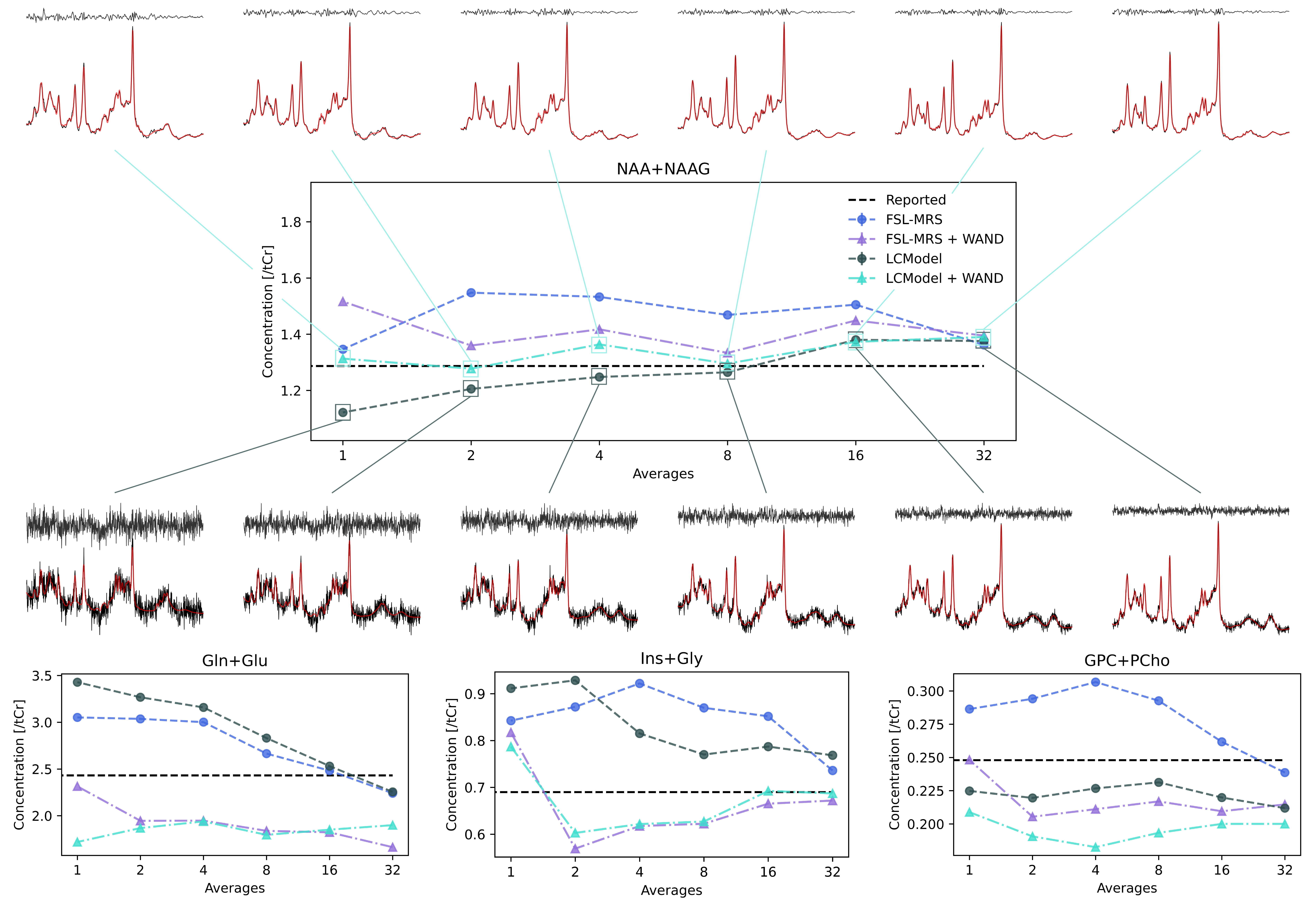}
    \caption{Shows the predicted metabolite concentration estimates of dataset 3 for varying NSA values using FSL-MRS and LCModel, with and without \ac{wand}, for the major metabolites. Additionally, respective LCModel fits are displaced, showcasing the representative spectra, the model fit, and the residuals.}
    \label{fig:invivo_noise_concs}
\end{figure*}

To analyze the robustness of the \ac{wand} method to in-vivo \ac{mrs} data, we utilize the individual averages of the 15 baseline acquisitions of the "fMRS in pain" data. Specifically, we average the datasets for different \ac{nsa} to obtain spectra of varying quality. Figure~\ref{fig:invivo_noise_concs} shows, for dataset 3, the influence of noise interference on the estimated metabolite concentrations for FSL-MRS and LCModel, with or without \ac{wand} processed spectra, for the major metabolites. Additionally, the metabolite concentration estimates reported in the work of Archibald et al.~\cite{Archibald2020MetaboliteAI} for 32 averages, obtained by LCModel, are included. The figure also presents the representative spectra, LCModel fits, and corresponding residuals for the different \ac{nsa}, highlighting the influence of noise with or without the use of \ac{wand}.

\begin{figure*}
    \centering
    \includegraphics[width=2\columnwidth, trim= 0 10mm 0 10mm, clip]{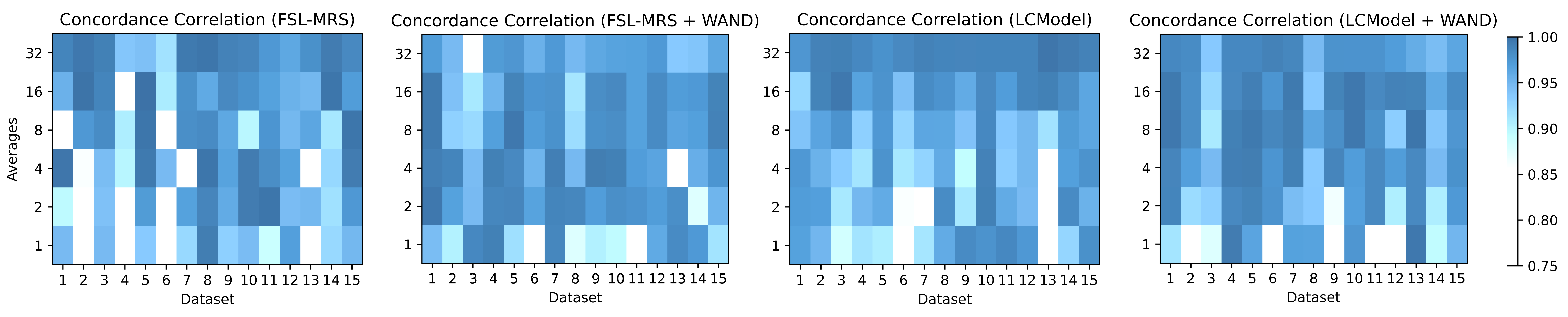}
    \caption{The obtained concordance correlation coefficients between the major metabolites reported  \cite{Archibald2020MetaboliteAI} and the obtained estimates of FLS-MRS and LCModel alone as well as in combination with \ac{wand} for varying \ac{nsa}.}

    \label{fig:invivo_noise_corrs}
\end{figure*}

To further assess the reproducibility of the metabolite estimates with varying \ac{nsa}, we computed the concordance correlation coefficients \cite{Lin1989CCC} between the reported major metabolite estimates for 32 averages and the obtained estimates for lower \ac{nsa} using different methods. Figure~\ref{fig:invivo_noise_corrs} presents the obtained correlations for FSL-MRS and LCModel alone, as well as in combination with \ac{wand}. It should be noted that the compared, reported values are LCModel estimates with a slightly different basis set. Therefore, we observe a strong correlation of nearly 1 at 32 averages for LCModel. In general, estimates when \acp{lcm} are combined with \ac{wand} show a stronger correlation throughout the lower \acp{nsa}.

\subsection{Lipid Contamination} \label{sssec:lipids}
\begin{figure*}
    \centering
    \includegraphics[width=2\columnwidth]{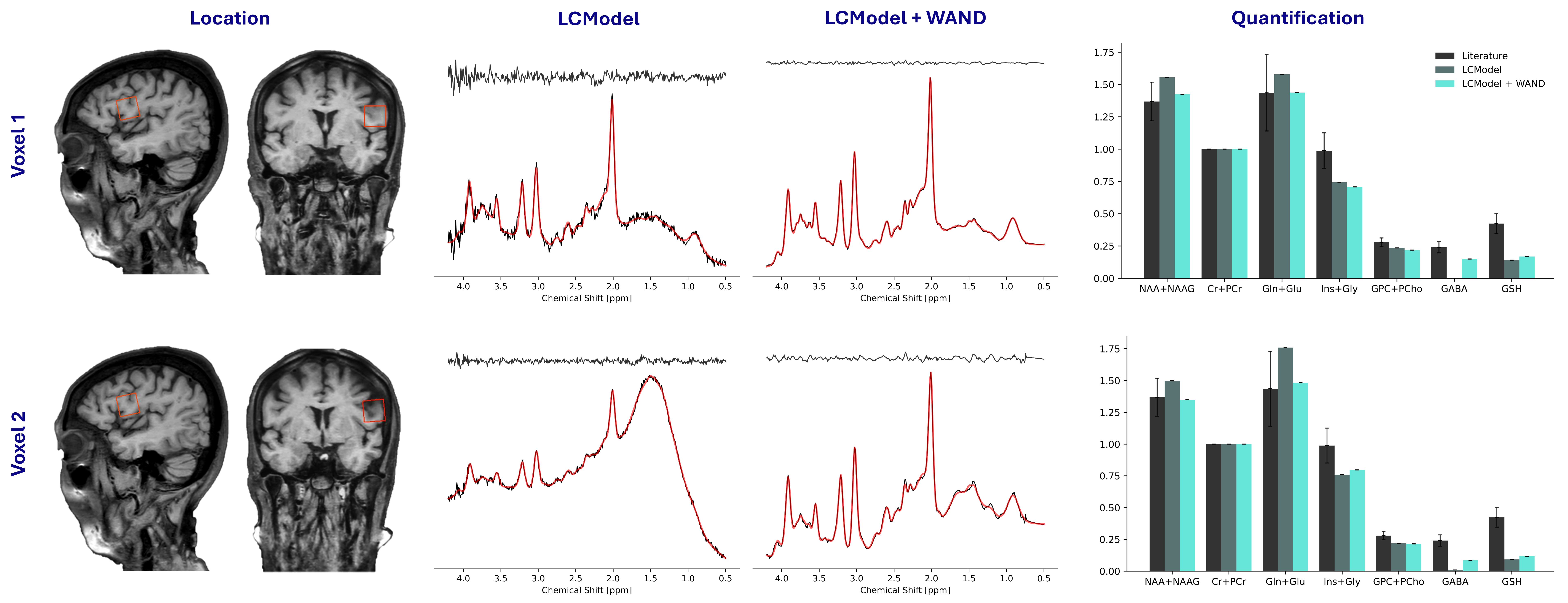}
    \caption{Acquired spectra, fitting, and quantification for two different voxel locations near the skull, with and without the use of \ac{wand} for processing. Voxel~1 exhibits minimal influence from lipid signals originating outside the brain, whereas Voxel~2 is significantly affected by these artifacts. }
    \label{fig:invivo_lipids}
\end{figure*}

In this section, we examine the capability of \ac{wand} to eliminate artifactual lipid signals originating from regions outside the brain \cite{Tk2020WaterAL}. To achieve this, we first acquire a reference spectrum from a voxel near the skull. We then slightly adjust the voxel’s position closer to the skull to obtain a spectrum from nearly the same voxel, but with subcutaneous fat contamination. Figure~\ref{fig:invivo_lipids} illustrates the exact voxel locations, the acquired spectra (with and without artifact removal using \ac{wand}), the LCModel fits, and the respective metabolite quantifications. These quantifications highlight the difference between modeling the lipid signals with basis functions and removing them using \ac{wand}. Additionally, we utilized the open-source database for in-vivo brain \ac{mrs} \cite{Gudmundson2023MetaanalysisAO} to identify relevant literature values for metabolite concentrations of that brain region. The database was queried for healthy/control literature and filtered by subject age (over 18), TE of 30 ms, \ac{press} sequence, proton \ac{mrs} only, left brain side, and voxel location to minimize variability. This filtering resulted in the following literature \cite{Burger2018TheImpactOfAcute, Burger2020TheRelationshipBetween, Smesny2018PrefrontalGE, Smesny2021AlterationsOfNeuro, Wang2021MetabolicAO}, from which metabolite concentrations were averaged, and a mean standard deviation was calculated. We notice a reasonable agreement with the literature for both methods, with slightly more constant concentration estimates obtained by applying the \ac{wand}. However, the experiment is limited by only one sample, tissue composition, the influence of movement and displacement, processing of the spectra, as well as the analysis itself. 

%artifactual lipid signals originating from regions outside of the brain, such as subcutaneous fat, \cite{Tk2020WaterAL} termed extracranial lipids \cite{kreis_terminology_2021}

%%%%%%%%%%%%%%%%%%
%%% Discussion %%%
%%%%%%%%%%%%%%%%%%
\section{Discussion}\label{sec:discussion}

The proposed \ac{wand} delivers an approach to adaptively categorize and separate wavelet coefficients with soft masks to account for overlapping signal structures. In contrast to traditional wavelet analysis methods for signal enhancement, such as \ac{sift} \cite{Doyle1994SIFTAP}, \ac{hsvd} \cite{Rowland2021ACO}, or wavelet thresholding \cite{Cancino2002SignalDN} that separate signal and noise components in a hard manner. Furthermore, by creating an artifact mask from the residual of the measured signal signal and all known signal components the artifact channel is not predicted but inferred by predicting all other signal components. This allows \ac{wand} to recover unpredictable random structures, such as Gaussian noise or signals generated by random walks.

%This study introduces \ac{wand}, a novel data-driven method for decomposing \ac{mrs} signals into metabolite-specific, baseline, and artifact components. 
The method demonstrates accurate decomposition in both simulated and in-vivo datasets, leading to significant improvements in quantification accuracy when combined with existing fitting methods like LCModel and FSL-MRS, as demonstrated in Figure~\ref{fig:synth_art_removal}. This is particularly evident in the case of spectra with strong artifacts, such as residual water signals in the 2016 MRS Fitting Challenge dataset or the lipid contaminations of Section~\ref{sssec:lipids}, where \ac{wand} preprocessing leads to refined concentration estimates.

While \ac{wand}'s data-driven nature allows for adaptability, it can introduce biases, and relying solely on simulations during the training process may limit generalization capabilities. The accuracy of the decomposition heavily relies on the comprehensiveness of the simulated data and the distributions of the model parameters summarized in Table~\ref{tab:sim_params}. Furthermore, the approach is trained for specific sequence parameters and is therefore dependent on these parameters to operate reliably, which is common for data-driven methods \cite{Sande2023AReviewOfML}. However, \ac{wand} offers significant advantages in terms of interpretability and reliability compared to conventional black-box deep learning methods. The decomposition of the \ac{mrs} signal into its individual components allows for visual inspection, enabling users to directly assess the quality of artifact removal and the isolation of metabolite signals (Figures~\ref{fig:synth_overview},~\ref{fig:synth_decomps},~\ref{fig:synth_art_removal}).

The combination of \ac{wand} with least-squares fitting methods might not be optimal. As \ac{wand} effectively removes Gaussian noise within the artifact component, the residual signal might no longer adhere to the Gaussian noise assumption inherent in least-squares methods. This could potentially impact the accuracy of quantification and might necessitate the exploration of alternative fitting approaches better suited for non-Gaussian residual noise.

Several areas for future research are important to consider.
\Ac{wand}'s ability to identify and remove consistent baseline effects and artifacts across multiple measurements makes it particularly well-suited for \ac{fmrs} applications. By applying the same baseline/artifact removal strategy to all time series data, \ac{wand} can improve the accuracy of dynamic metabolite concentration estimates while only introducing potential systemic biases. Additionally, tailoring \ac{wand} to specific applications, such as focusing on clinically relevant metabolites or specific types of artifacts, could further enhance its performance. This could be achieved by implementing a weighted loss function that prioritizes the accurate decomposition of specific metabolites or artifact patterns.

%While the Morlet wavelet was chosen for its computational efficiency and ability to separate metabolites from noise and baseline variations, the selection of an optimal wavelet involves inherent trade-offs. 
The choice of wavelet significantly impacts the decomposition process. Depending on the length, sparsity, and abruptness of signal changes, different wavelets are more suitable for distinguishing between signal and noise in the wavelet domain. \cite{Sahoo2024OptimalWaveletSel} In the case of \ac{wand}, we encounter numerous components with varying frequency characteristics and different mean sparsity changes, making the selection of a single optimal wavelet impractical. Additionally, the artifact channel itself is highly variable, containing only noise at times, but also exhibiting smooth baseline signal changes and potential sharp water and lipid peak contaminations. Therefore, we compromise by selecting the computationally efficient Morlet wavelet as it has shown to be less sensitive to higher-order noise as well as baseline variation which allows for a more accurate separation of the metabolite signals from contaminations, and while preserving the integrity of the signals. \cite{Suvichakorn2008MorletWavelet, Suvichakorn2008QuantificationMeth}
Different wavelets may be more suitable depending on the specific characteristics of the signal being analyzed. Exploring adaptive wavelet methods \cite{Shao2021ModifiedSA} could lead to more robust and adaptable artifact removal strategies tailored to the specific features of individual MRS datasets.

Furthermore, investigating the potential of incorporating \ac{wand}'s decomposed signal components as a basis set for quantification or augmenting existing basis sets with the artifact channel could be a promising direction for future research.

%%%%%%%%%%%%%%%%%%
%%% Conclusion %%%
%%%%%%%%%%%%%%%%%%
\section{Conclusion}\label{sec:conclusion}

This paper introduced \ac{wand}, a novel data-driven method for decomposing \ac{mrs} signals into metabolite-specific, baseline, and artifact components. The method takes advantage of the enhanced separability of these components within the wavelet domain, using a U-Net architecture to predict soft masks for wavelet coefficients. These masks are then used to isolate and reconstruct individual signal components.
\Ac{wand} demonstrated accurate decomposition performance on simulated spectra, as evidenced by the low \ac{mse} values reported in Table~\ref{tab:decomp} and the visualizations in Figures~\ref{fig:synth_overview} and~\ref{fig:synth_decomps}. Notably, 
\ac{wand}'s ability to infer the artifact component by predicting all other signal components allows for the capture of unpredictable, random structures without explicit artifact modeling.
The application of \ac{wand} to both simulated and in-vivo data highlighted its potential to enhance the accuracy of metabolite quantification. By removing artifacts, \ac{wand} improves the performance of linear combination model fitting, as shown in Figures~\ref{fig:synth_art_removal},~\ref{fig:synth_comparison},~\ref{fig:challenge_conc}, and~\ref{fig:invivo_lipids}. The method's robustness is further supported by the improved concordance correlation coefficients observed when \ac{wand} is combined with LCM methods for in-vivo data with varying noise levels (Figure~\ref{fig:invivo_noise_corrs}).

\section*{Acknowledgments}
The authors would like to thank Jessica Archibald and co-authors of the work \cite{Archibald2020MetaboliteAI} for making their data publicly available and for providing their LCModel quantifications.
The authors would also like to extend their gratitude to Maarten Versluis for their valuable contributions to the acquisition of the in-vivo data presented in Section~\ref{sssec:lipids}. Additionally, the authors thank Philips for the use of their facilities and support in acquiring the in-vivo data.
The authors would also like to thank Kay Igwe for their help with the MARSS simulations.
This work was in part funded by Spectralligence (EUREKA IA Call, ITEA4 project 20209) and NWO VIDI (VI.Vidi.223.085).

% \subsection*{Author contributions}
% This is an author contribution text. This is an author contribution text. This is an author contribution text. This is an author contribution text. This is an author contribution text. This is an author contribution text. 

% \subsection*{Financial disclosure}
% None reported.

% \subsection*{Conflict of interest}
% The authors declare no potential conflict of interests.

\section*{Data Availability Statement}
The source code used in our experiments for the method, data simulation, and analysis can be found at \url{https://github.com/julianmer/WAND-for-MRS}.

\bibliography{refs/refs, refs/lr_refs, refs/my_refs}
% \vfill\pagebreak

%\section*{Supporting information}
%...

\appendix
The Appendix of this paper is composed of three distinct sections: \ref{app:add_materials} Additional Materials, \ref{app:details} Implementation Details, and \ref{app:mrs_in_mrs} Minimum Reporting Standards in MRS. Each section contains supplementary information that aims to further enhance the workings and findings of the main text.

\section{Additional Materials} \label{app:add_materials}
This appendix provides supplementary materials and analyses that support the findings and discussions presented in the main text.

\begin{figure*}
    \centering
    \includegraphics[width=1.7\columnwidth]{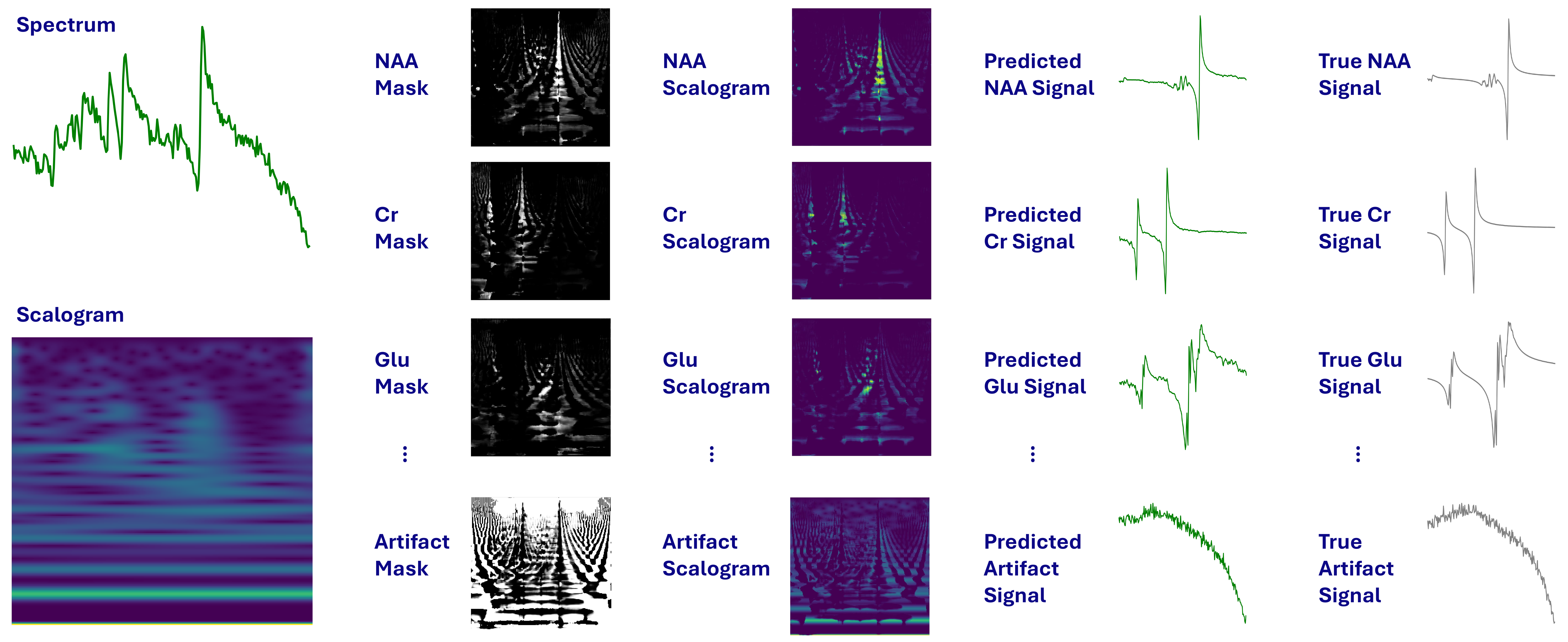}
    \caption{An illustrative summary of the steps in the \ac{wand}, showcasing the imaginary part of the spectrum presented in Figure~\ref{fig:synth_overview}, its associated scalogram, several predicted masks, their respective applied scalograms, and the ultimately obtained spectral components along with the corresponding ground truth signals.} \label{fig:synth_overview_imag}
\end{figure*}

\begin{figure*}
    \centering
    \includegraphics[width=1.9\columnwidth]{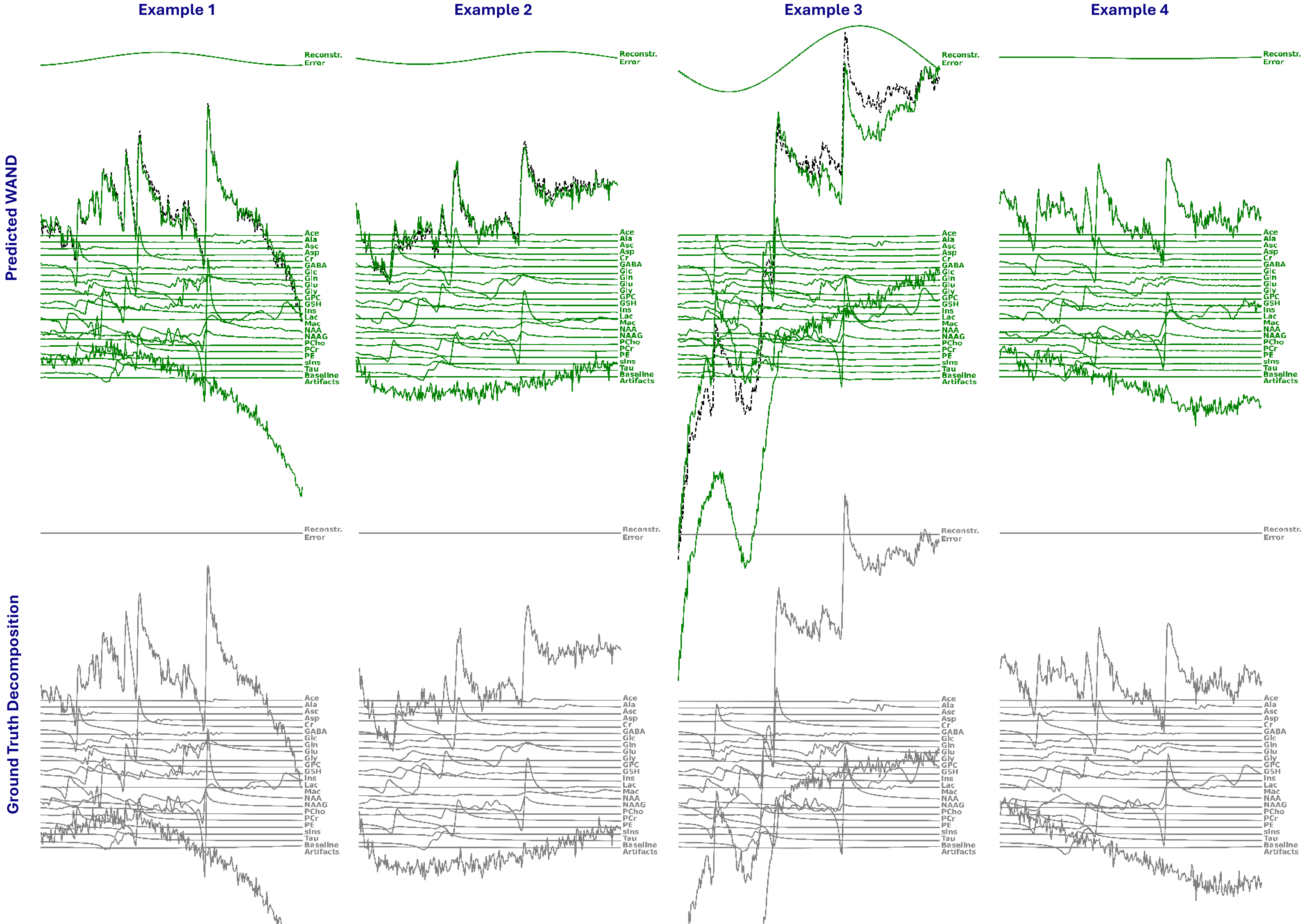}
    \caption{Shows the imaginary parts of the complete predicted decomposition for the four selected examples of Figure~\ref{fig:synth_decomps}, alongside their corresponding ground truth decompositions in the range of 0.5 to 4.5 ppm.}
    \label{fig:synth_decomps_imag}
\end{figure*}

\begin{figure*}
    \centering
    \includegraphics[width=1.75\columnwidth, trim= 0 10mm 0 1mm, clip]{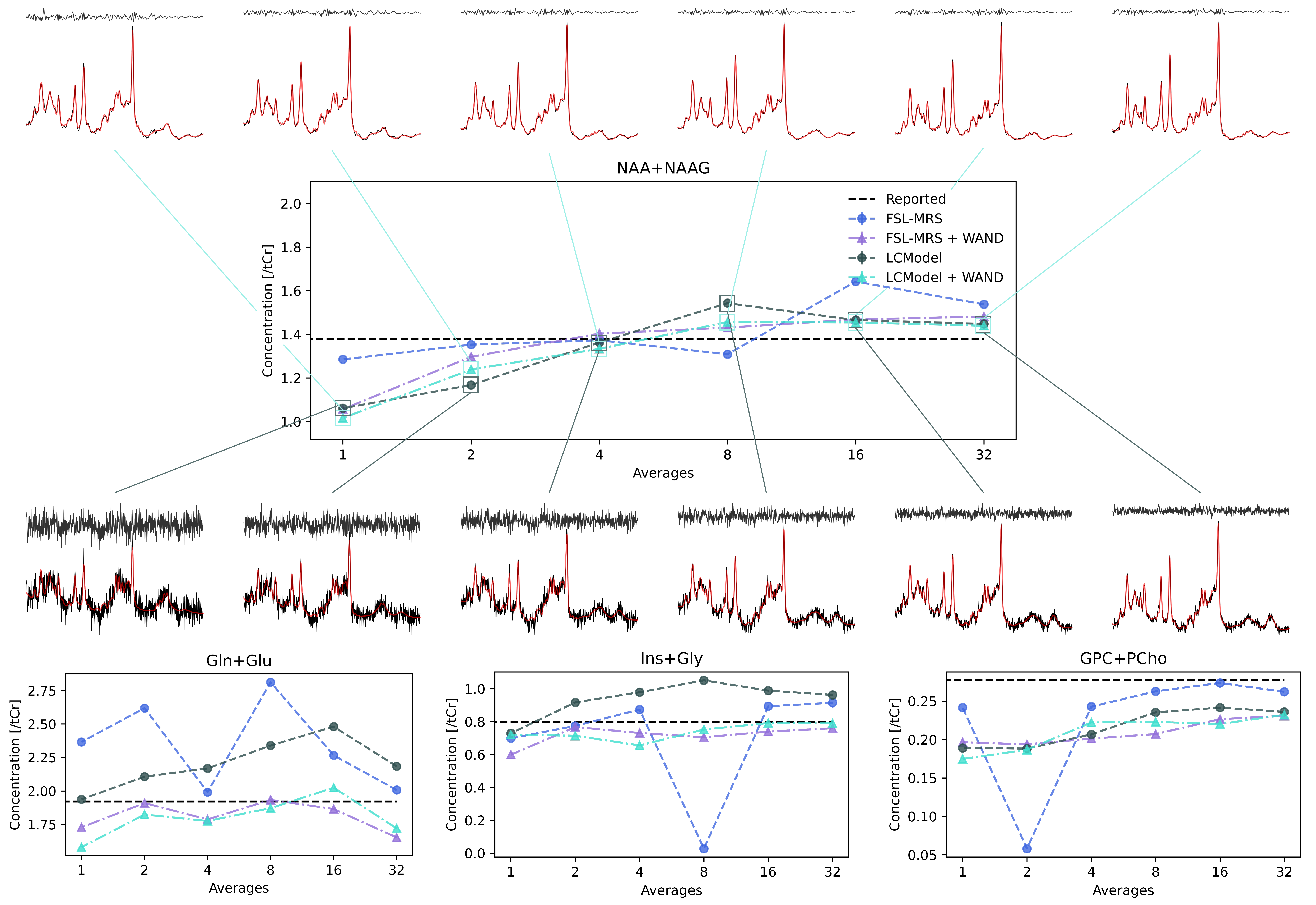}
    \caption{Shows the predicted metabolite concentration estimates of dataset 1 for varying NSA values using FSL-MRS and LCModel, with and without \ac{wand}, for the major metabolites.}
    \label{fig:invivo_noise_concs0}
\end{figure*}

\begin{figure*}
    \centering
    \includegraphics[width=1.75\columnwidth, trim= 0 10mm 0 1mm, clip]{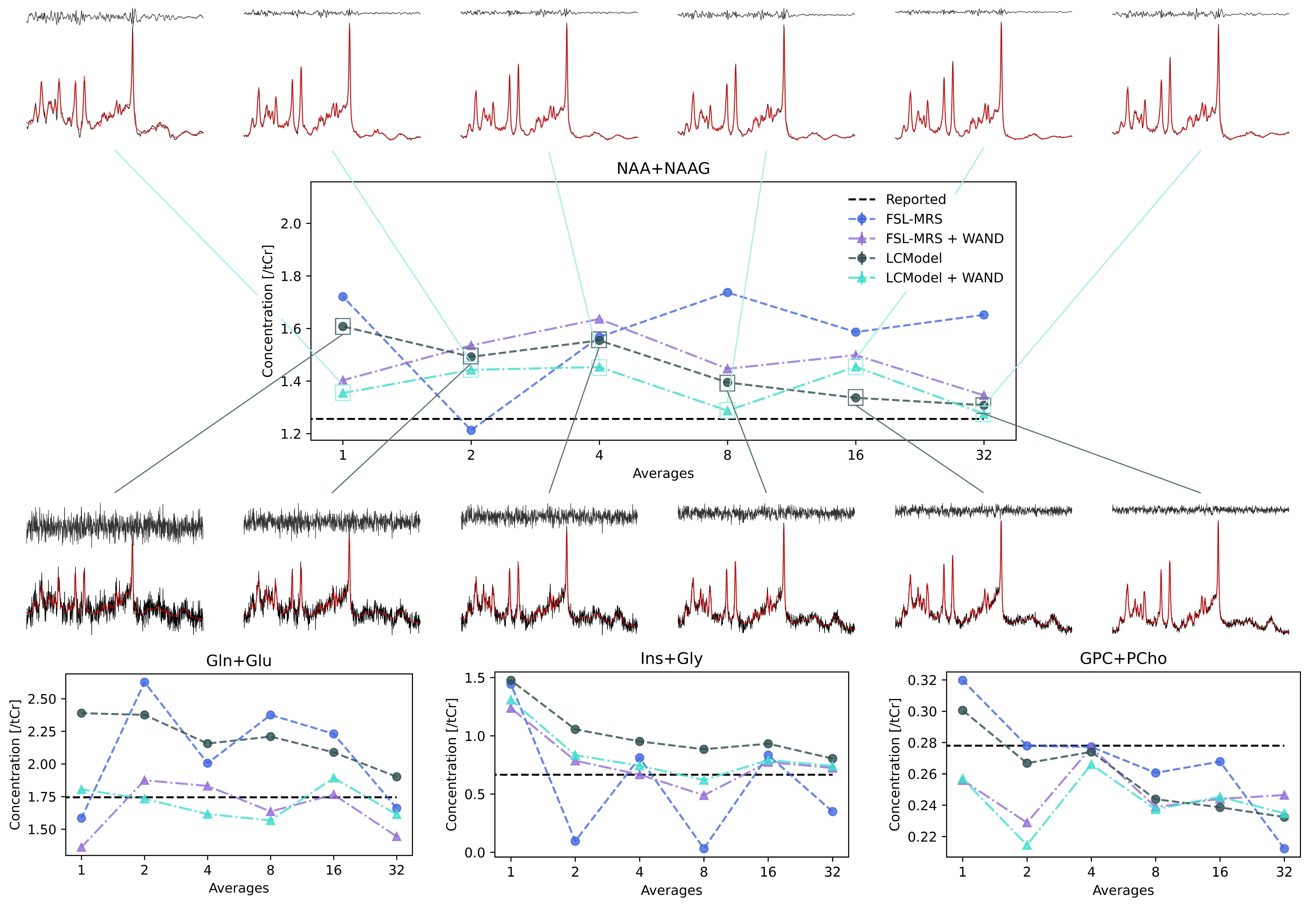}
    \caption{Shows the predicted metabolite concentration estimates of dataset 6 for varying NSA values using FSL-MRS and LCModel, with and without \ac{wand}, for the major metabolites.}
    \label{fig:invivo_noise_concs5}
\end{figure*}

\begin{figure*}
    \centering
    \includegraphics[width=1.75\columnwidth, trim= 0 10mm 0 1mm, clip]{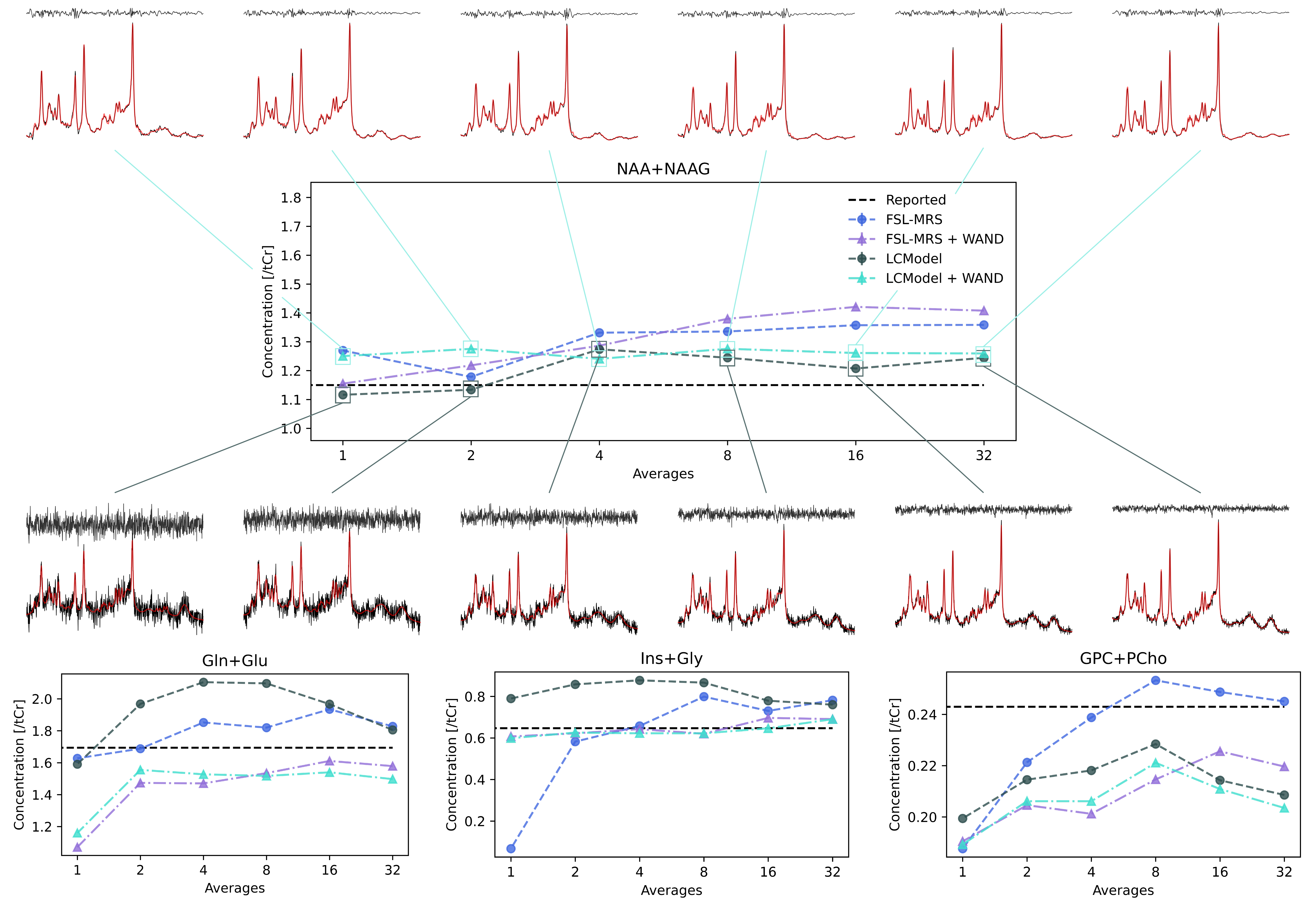}
    \caption{Shows the predicted metabolite concentration estimates of dataset 11 for varying NSA values using FSL-MRS and LCModel, with and without \ac{wand}, for the major metabolites.}
    \label{fig:invivo_noise_concs10}
\end{figure*}

\begin{itemize}
    \item Figure~\ref{fig:synth_overview_imag} and Figure~\ref{fig:synth_decomps_imag} present visualizations analogous to those shown in the main text (Figures~\ref{fig:synth_overview} and~\ref{fig:synth_decomps}), but focusing on the imaginary components of the \ac{wand} framework.
    
    \item Table~\ref{tab:decomp_all} expands upon the error analysis presented in Table~\ref{tab:decomp} by providing a wider range of error metrics for the predicted signal decompositions.
    
    \item Figures\ref{fig:invivo_noise_concs0},~\ref{fig:invivo_noise_concs5}, and~\ref{fig:invivo_noise_concs10} show metabolite concentration estimates for datasets 0, 5 and 10, analogous to Figure~\ref{fig:invivo_noise_concs} which shows the same for dataset 3.
\end{itemize}
\pagebreak

\begin{table*}[b]
    % \centering
    \caption{Unscaled \Ac{mse}, individually scaled component \ac{mse} (each normalized to an absolute value of 1), and cosine similarity, as well as the \ac{se} for all those measures (given in brackets with ±), of the real and imaginary parts of the predicted and ground truth signal components, computed over a test set of 1000 random samples.} \label{tab:decomp_all}%
    \vspace{2mm}
    \begin{tabular*}{2\columnwidth}{@{\extracolsep{\fill}\hspace{1mm}}lllllll@{\extracolsep{\fill}\hspace{1mm}}}
    \toprule
        \textbf{Component} & \textbf{Un. MSE Real} & \textbf{Ind. MSE Real} & \textbf{Cos Sim. Real} & \textbf{Un. MSE Imag} & \textbf{Ind. MSE Imag} & \textbf{Cos Sim. Imag} \\
    \midrule
        Ace & 1.95e+3\ (± 1.9e+2) & 7.11e-2\ (± 3.4e-3) & 0.68\ (± 1.1e-2) & 2.20e+3\ (± 2.0e+2) & 4.99e-2\ (± 1.2e-3) & 0.55\ (± 9.0e-3) \\ 
        Ala & 2.13e+3\ (± 2.0e+2) & 4.49e-2\ (± 2.6e-3) & 0.78\ (± 9.4e-3) & 2.41e+3\ (± 2.0e+2) & 3.14e-2\ (± 1.0e-3) & 0.72\ (± 8.4e-3) \\ 
        Asc & 2.08e+3\ (± 1.9e+2) & 4.56e-2\ (± 2.6e-3) & 0.82\ (± 8.3e-3) & 2.38e+3\ (± 2.1e+2) & 2.76e-2\ (± 9.9e-4) & 0.76\ (± 8.4e-3) \\ 
        Asp & 1.95e+3\ (± 1.9e+2) & 6.37e-2\ (± 3.0e-3) & 0.73\ (± 1.0e-2) & 2.21e+3\ (± 2.0e+2) & 3.46e-2\ (± 1.3e-3) & 0.63\ (± 9.7e-3) \\ 
        Cr & 6.44e+3\ (± 3.1e+2) & 6.63e-4\ (± 4.7e-5) & 0.99\ (± 4.8e-4) & 7.37e+3\ (± 3.4e+2) & 8.88e-4\ (± 6.3e-5) & 0.99\ (± 5.5e-4 \\ 
        GABA & 2.09e+3\ (± 1.9e+2) & 4.58e-2\ (± 2.4e-3) & 0.74\ (± 9.7e-3) & 2.26e+3\ (± 1.9e+2) & 3.15e-2\ (± 1.4e-3) & 0.74\ (± 1.1e-2) \\ 
        Glc & 2.26e+3\ (± 2.0e+2) & 2.63e-2\ (± 1.8e-3) & 0.89\ (± 6.8e-3) & 2.63e+3\ (± 2.0e+2) & 2.02e-2\ (± 8.8e-4) & 0.86\ (± 6.6e-3) \\ 
        Gln & 2.86e+3\ (± 1.9e+2) & 1.44e-2\ (± 9.5e-4) & 0.93\ (± 4.1e-3) & 3.23e+3\ (± 2.1e+2) & 8.19e-3\ (± 4.7e-4) & 0.93\ (± 4.3e-3) \\ 
        Glu & 4.69e+3\ (± 2.6e+2) & 4.02e-3\ (± 2.9e-4) & 0.98\ (± 1.6e-3) & 5.10e+3\ (± 2.5e+2) & 2.73e-3\ (± 1.7e-4) & 0.98\ (± 1.2e-3) \\ 
        Gly & 2.00e+3\ (± 1.9e+2) & 5.21e-2\ (± 2.9e-3) & 0.77\ (± 9.5e-3) & 2.23e+3\ (± 2.0e+2) & 2.88e-2\ (± 1.0e-3) & 0.73\ (± 8.2e-3) \\ 
        GPC & 4.16e+3\ (± 2.3e+2) & 2.71e-3\ (± 2.5e-4) & 0.97\ (± 1.9e-3) & 4.42e+3\ (± 2.3e+2) & 2.97e-3\ (± 2.8e-4) & 0.97\ (± 2.1e-3) \\ 
        GSH & 2.68e+3\ (± 1.9e+2) & 7.31e-3\ (± 6.3e-4) & 0.94\ (± 3.6e-3) & 3.37e+3\ (± 2.1e+2) & 8.82e-3\ (± 5.7e-4) & 0.93\ (± 3.8e-3) \\ 
        Ins & 4.70e+3\ (± 2.4e+2) & 1.87e-3\ (± 1.2e-4) & 0.98\ (± 8.7e-4) & 4.87e+3\ (± 2.7e+2) & 2.90e-3\ (± 2.2e-4) & 0.98\ (± 1.1e-3) \\ 
        Lac & 2.01e+3\ (± 1.9e+2) & 6.45e-2\ (± 3.0e-3) & 0.69\ (± 1.0e-2) & 2.28e+3\ (± 2.0e+2) & 4.58e-2\ (± 1.2e-3) & 0.62\ (± 9.3e-3) \\ 
        Mac & 1.04e+4\ (± 5.9e+2) & 6.64e-3\ (± 4.8e-4) & 0.97\ (± 2.4e-3) & 1.47e+4\ (± 8.3e+2) & 3.31e-3\ (± 2.3e-4) & 0.99\ (± 1.0e-3) \\ 
        NAA & 5.59e+3\ (± 2.7e+2) & 3.75e-4\ (± 2.6e-5) & 0.99\ (± 3.7e-4) & 6.41e+3\ (± 3.1e+2) & 7.95e-4\ (± 1.0e-4) & 0.99\ (± 3.9e-4 \\ 
        NAAG & 3.57e+3\ (± 2.1e+2) & 9.60e-3\ (± 9.0e-4) & 0.92\ (± 4.4e-3) & 3.76e+3\ (± 2.1e+2) & 8.70e-3\ (± 5.9e-4) & 0.93\ (± 3.9e-3) \\ 
        PCho & 3.43e+3\ (± 2.0e+2) & 6.92e-3\ (± 7.0e-4) & 0.94\ (± 3.9e-3) & 3.83e+3\ (± 2.2e+2) & 6.90e-3\ (± 5.0e-4) & 0.93\ (± 3.9e-3) \\ 
        PCr & 5.12e+3\ (± 2.4e+2) & 1.39e-3\ (± 1.0e-4) & 0.98\ (± 9.2e-4) & 6.12e+3\ (± 2.5e+2) & 1.71e-3\ (± 1.3e-4) & 0.98\ (± 1.1e-3) \\ 
        PE & 2.10e+3\ (± 1.9e+2) & 4.19e-2\ (± 2.3e-3) & 0.82\ (± 8.0e-3) & 2.35e+3\ (± 2.0e+2) & 2.77e-2\ (± 1.0e-3) & 0.79\ (± 8.1e-3) \\ 
        sIns & 2.02e+3\ (± 1.9e+2) & 2.65e-2\ (± 2.1e-3) & 0.86\ (± 7.2e-3) & 2.32e+3\ (± 2.0e+2) & 1.71e-2\ (± 8.7e-4) & 0.83\ (± 7.1e-3) \\ 
        Tau & 3.66e+3\ (± 2.2e+2) & 7.07e-3\ (± 6.5e-4) & 0.97\ (± 2.1e-3) & 4.20e+3\ (± 2.3e+2) & 4.78e-3\ (± 3.7e-4) & 0.97\ (± 2.0e-3) \\ 
        Baseline & 2.95e+3\ (± 1.9e+2) & 1.11e-1\ (± 2.9e-3) & 0.37\ (± 1.7e-2) & 3.07e+3\ (± 2.1e+2) & 1.52e-1\ (± 3.1e-3) & 0.19\ (± 1.8e-2) \\ 
        Artifacts & 1.05e+5\ (± 1.3e+4) & 1.91e-3\ (± 7.6e-5) & 0.98\ (± 7.7e-4) & 1.35e+5\ (± 1.1e+4) & 2.22e-3\ (± 7.1e-5) & 0.98\ (± 6.6e-4) \\ 
    \bottomrule
    \end{tabular*}
\end{table*}

\section{Implementation Details} \label{app:details}

This section offers a closer look at the specific implementation choices made in developing the \ac{wand} method.

\begin{itemize}
    \item Table \ref{tab:unet} dissects the U-Net architecture employed in \ac{wand}, outlining the specific details of each layer within the network.
    
    \item Table \ref{tab:ml_params} summarizes the settings and training parameters of the \ac{wand} framework: This table serves as a quick reference guide for the various settings and hyperparameters used during the training and operation of the WAND framework. %This detailed breakdown of parameters enables reproducibility.
\end{itemize}

\begin{table*}[b]%
    \caption{Specific neural network layer details of the U-Net architecture of \ac{wand} implemented in PyTorch. \hspace{12cm}} \label{tab:unet}%
    % \caption{U-Net architecture details.}
    \vspace{2mm}
    \begin{tabular*}{2\columnwidth}{@{\extracolsep{\fill}\hspace{1mm}}ccccccc@{\extracolsep{\fill}\hspace{1mm}}}
    \toprule
    \textbf{Layer}$^{*}$  & \textbf{Operation} & \textbf{Kernel Size} & \textbf{Stride} & \textbf{Padding} & \textbf{In Channels} & \textbf{Out Channels} \\
    \midrule
    Input & - & - & - & - & n\_channels & - \\
    \multirow{2}{*}{DoubleConv} & Conv2d, BN, ReLU & 3 & 1 & 1 & \multirow{2}{*}{n\_channels} & \multirow{2}{*}{64} \\
    & Conv2d, BN, ReLU & 3 & 1 & 1 \\
    \multirow{2}{*}{Down} & MaxPool2d & 2 & 2 & 0 & \multirow{2}{*}{64} & \multirow{2}{*}{128} \\
    & DoubleConv & 3 & 1 & 1 \\
    \multirow{2}{*}{Down} & MaxPool2d & 2 & 2 & 0 & \multirow{2}{*}{128} & \multirow{2}{*}{256} \\
    & DoubleConv & 3 & 1 & 1 \\
    \multirow{2}{*}{Down} & MaxPool2d & 2 & 2 & 0 & \multirow{2}{*}{256} & \multirow{2}{*}{512} \\
    & DoubleConv & 3 & 1 & 1 \\
    \multirow{2}{*}{Down} & MaxPool2d & 2 & 2 & 0 & \multirow{2}{*}{512} & \multirow{2}{*}{1024} \\
    & DoubleConv & 3 & 1 & 1 \\
    \multirow{2}{*}{Up} & ConvTranspose2d & 2 & 2 & 0 & \multirow{2}{*}{1024} & \multirow{2}{*}{512} \\
    & DoubleConv & 3 & 1 & 1 \\
    \multirow{2}{*}{Up} & ConvTranspose2d & 2 & 2 & 0 & \multirow{2}{*}{512} & \multirow{2}{*}{256} \\
    & DoubleConv & 3 & 1 & 1 \\
    \multirow{2}{*}{Up} & ConvTranspose2d & 2 & 2 & 0 & \multirow{2}{*}{256} & \multirow{2}{*}{128} \\
    & DoubleConv & 3 & 1 & 1 \\
    \multirow{2}{*}{Up} & ConvTranspose2d & 2 & 2 & 0 & \multirow{2}{*}{128} & \multirow{2}{*}{64} \\
    & DoubleConv & 3 & 1 & 1 \\
    OutConv & Conv2d & 1 & 1 & 0 & 64 & n\_classes \\
    \bottomrule
    \end{tabular*}
    \begin{tablenotes}%%[341pt]
    \item[$^{*}$] Dropout is added in between each of the listed layers.
\end{tablenotes}
\end{table*}

\begin{table*}[b]%
    \caption{Overview of the settings and training parameters of the \ac{wand} framework. \hspace{12cm}} \label{tab:ml_params}%
    \vspace{2mm}
    \begin{tabular*}{2\columnwidth}{@{\extracolsep{\fill}\hspace{1mm}}lll@{\extracolsep{\fill}\hspace{1mm}}}
    \toprule
    \textbf{Parameter} & \textbf{Value} & \textbf{Description} \\
    \midrule
    \textbf{Data} \\
    dataType & norm\_rw\_p & The preset type of data to be simulated for training. \\
    basisFmt &  & The preset basis format, defaults to empty string. \\
    path2basis & path & Path to the desired basis set. \\
    specType & synth & Type of the spectra, i.e. selects preset ppm region. \\
    ppmlim & (0.5, 4.5) & The ppm limits for the spectra (used if specType='auto'). \\
    \midrule
    \textbf{Architecture} \\
    arch & unet & The neural network architecture. \\
    loss & mse & The loss function to train the model, default Mean Squared Error (MSE). \\
    model & wand & Model type. \\
    \midrule
    \textbf{Other} \\
    load\_model & False & Load a pre-trained model. \\
    path2trained &  & Path to a trained model (if load\_model=True). \\
    skip\_train & False & Skip training (for loading and testing purposes). \\
    \midrule
    \textbf{Training Parameters} \\
    batch & 16 & The batch size. \\
    trueBatch & 16 & Accumulates the gradients over trueBatch/batch. \\
    check\_val\_every\_n\_epoch & None & None, if trained with generator, otherwise the number of epochs between validations. \\
    dropout & 0 & The proportion of nodes to drop out. \\
    l1\_reg & 0 & L1 regularization added to the loss. \\
    l2\_reg & 0 & L2 regularization added to the loss. \\
    learning & 0.0001 & Learning rate. \\
    max\_epochs & -1 & Maximum number of epochs. \\
    max\_steps & -1 & Maximum number of steps/iterations. \\
    val\_check\_interval & 256 & None, if fixed training set, otherwise the number of iterations per between validations. \\
    val\_size & 1,024 & Validation size (in samples). \\
    \bottomrule
    \end{tabular*}
\end{table*}

\section{MRS in MRS} \label{app:mrs_in_mrs}

This section focuses on ensuring transparency and reproducibility by adhering to the \ac{mrsinmrs} guidelines \cite{lin_minimum_2021}. It presents tables that detail the acquisitions, experimental setup, processing, and data analysis methods employed in the study.

\begin{table*}[b]%
    \caption{Minimum Reporting Standards for in-vivo Magnetic Resonance Spectroscopy (MRSinMRS) \cite{lin_minimum_2021} for the data of Section~\ref{sssec:noise}. \hspace{12cm}} \label{tab:mrsinmrs_frms}%
    % \caption{U-Net architecture details.}
    \vspace{2mm}
    \begin{tabular*}{2\columnwidth}{@{\extracolsep{\fill}\hspace{1mm}}ll@{\extracolsep{\fill}\hspace{1mm}}}
    \toprule
    \textbf{Site (name or number)} & \textbf{See work of Archibald et al. \cite{Archibald2020MetaboliteAI}} \\
    \midrule
    \textbf{1. Hardware} & \\
    a. Field strength [T] & 3 T (127.795 MHz) \\
    b. Manufacturer & Philips \\
    c. Model (software version if available) &  \\
    d. RF coils: nuclei (transmit/receive), number of channels, type, body part & See work of Archibald et al. \cite{Archibald2020MetaboliteAI}  \\
    e. Additional hardware &  \\
    \midrule
    \textbf{2. Acquisition} & \\
    a. Pulse sequence & SV PRESS \\
    b. Volume of interest (VOI) locations & Anterior cingulate cortex \\
    c. Nominal VOI size [cm3, mm3] & 30 × 25 × 15 mm$^3$ \\
    d. Echo time (TE) / repetition time (TR) [ms, s] & 22 ms / 4 s \\
    e. Total number of excitations or acquisitions per spectrum & 32 averages \\
    f. Additional sequence parameters & 2000 Hz bandwidth, 2048 sample points, \\
    g. Water suppression method &  \\
    h. Shimming method, reference peak, and thresholds for "acceptance of shim" chosen &  See work of Archibald et al. \cite{Archibald2020MetaboliteAI}   \\
    i. Triggering or motion correction method (respiratory, peripheral, cardiac triggering) &  \\
    \midrule
    \textbf{3. Data Analysis Methods and Outputs} & \\
    a. Analysis software & In-house Python scripts, \\
    & FSL-MRS \cite{clarke_fslmrs_2021} (version 2.1.17), \\
    & LCModel \cite{provencher_estimation_1993} (version 6.3-1L) \\
    b. Processing steps (deviating from quoted reference or product) & NIfTI-MRS Header (ProcessingApplied): \\
    & \textit{from fsl\_mrs.utils.preproc import nifti\_mrs\_proc as proc} \\
    & Method: "RF coil combination", \\
    & Details: \textit{proc.coilcombine, reference=None,} \\
    & \hspace{1cm} \textit{no\_prewhiten=True}, \\
    & Method: "Frequency and phase correction", \\
    & Details: \textit{proc.align, dim=DIM\_DYN, window=None,} \\
    & \hspace{1cm } \textit{target=None, ppmlim=(0, 8), niter=2}, \\
    & Method: "Signal averaging", \\
    & Details: \textit{proc.average, dim=DIM\_DYN}, \\ 
    & Method: "Eddy current correction",  \\
    & Details: \textit{proc.ecc, reference=}, \\
    & Method: "Nuisance peak removal", \\
    & Details: \textit{proc.remove\_peaks, limits=[-0.15, 0.15],}\\
    & \hspace{1cm} \textit{limit\_units=ppm}, \\
    & Method: "Frequency and phase correction", \\
    & Details: \textit{proc.shift\_to\_reference, ppm\_ref=3.027,}\\
    & \hspace{1cm} \textit{peak\_search=(2.9, 3.1), use\_avg=False}, \\
    & Method: "Phasing", \\
    & Details: \textit{proc.phase\_correct, ppmlim=(2.9, 3.1),}\\
    & \hspace{1cm} \textit{ hlsvd=False, use\_avg=False}, \\
    c. Output measure (e.g. absolute concentration, institutional units, ratio) & Referenced to tCr \\
    d. Quantification references and assumptions, fitting model assumptions & Basis set simulated using MRSCloud \cite{Hui2022MRSCloud} \\
    & (metabolites from Table~\ref{tab:sim_params}, excluding Glc and MM) \\
    & LCModel control:\\
    & \hspace{1cm}\textit{\$LCMODL, nunfil=2048, deltat=0.0005,} \\
    & \hspace{1cm}\textit{hzpppm=127.731, ppmst=4.2, ppmend=0.5,} \\
    & \hspace{1cm}\textit{dows=F, doecc=F, neach=99, filbas='example.basis',} \\
    & \hspace{1cm}\textit{filraw='example.raw', filps='example.ps',} \\
    & \hspace{1cm}\textit{filcoo='example.coord', lcoord=9, nomit=2,} \\
    & \hspace{1cm}\textit{chomit(1)='-CrCH2', chomit(2)='CrCH2',} \\
    & \hspace{1cm}\textit{namrel='Cr+PCr', \$END} \\
    & FSL-MRS: \textit{from fsl\_mrs.utils import fitting} \\
    & \textit{fitting.fit\_FSLModel, method='Newton' ppmlim=(0.5, 4.2),} \\
    & \hspace{1cm}\textit{baseline\_order=4} \\
    \midrule
    \textbf{4. Data Quality} & \\
    a. Reported variables (SNR, linewidth (with reference peaks)) & S/N = 17 - 24, FWHM = 0.02 - 0.05 ppm (LCModel estimates)\\
    b. Data exclusion criteria & - \\
    c. Quality measures of postprocessing model fitting & - \\
    d. Sample spectrum & See Figures~\ref{fig:invivo_noise_concs}, \ref{fig:invivo_noise_concs0}, \ref{fig:invivo_noise_concs5}, and \ref{fig:invivo_noise_concs10}
    \end{tabular*}
\end{table*}

\begin{table*}[b]%
    \caption{MRSinMRS for the data of Section~\ref{sssec:lipids}. \hspace{12cm}} \label{tab:mrsinmrs_lipids}%
    % \caption{U-Net architecture details.}
    \vspace{2mm}
    \begin{tabular*}{2\columnwidth}{@{\extracolsep{\fill}\hspace{1mm}}ll@{\extracolsep{\fill}\hspace{1mm}}}
    \toprule
    \textbf{Site (name or number)} & \textbf{Philips, Best, The Netherlands} \\
    \midrule
    \textbf{1. Hardware} & \\
    a. Field strength [T] & 3 T (127.752504 MHz) \\
    b. Manufacturer & Philips \\
    c. Model (software version if available) & Ingenia Elition X 3.0T \\
    & Gyroscan SW release: 11.1-0, \\
    & Reconstruction Host SW version: 3, \\
    & Reconstruction AP SW version: 3 \\
    d. RF coils: nuclei (transmit/receive), number of channels, type, body part & 1H, 15 channel, head coil \\
    e. Additional hardware & - \\
    \midrule
    \textbf{2. Acquisition} & \\
    a. Pulse sequence & SV PRESS \\
    b. Volume of interest (VOI) locations & See Figure~\ref{fig:invivo_lipids} \\
    c. Nominal VOI size [cm3, mm3] & 20 × 20 × 20 mm$^3$ \\
    d. Echo time (TE) / repetition time (TR) [ms, s] & 30 ms / 4 s \\
    e. Total number of excitations or acquisitions per spectrum & 64 averages \\
    f. Additional sequence parameters & 4000 Hz bandwidth, 2048 sample points, \\
    g. Water suppression method & VAPOR \\
    h. Shimming method, reference peak, and thresholds for "acceptance of shim" chosen & PB-volume (pencil beam volume), 2nd order shimming \\
    i. Triggering or motion correction method (respiratory, peripheral, cardiac triggering) & - \\
    \midrule
    \textbf{3. Data Analysis Methods and Outputs} & \\
    a. Analysis software & In-house Python scripts, \\
    & FSL-MRS \cite{clarke_fslmrs_2021} (version 2.1.17), \\
    & LCModel \cite{provencher_estimation_1993} (version 6.3-1L) \\
    b. Processing steps (deviating from quoted reference or product) & NIfTI-MRS Header (ProcessingApplied): \\
    & \textit{from fsl\_mrs.utils.preproc import nifti\_mrs\_proc as proc} \\
    & Method: "Eddy current correction",  \\
    & Details: \textit{proc.ecc, reference=}, \\
    & Method: "Nuisance peak removal", \\
    & Details: \textit{proc.remove\_peaks, limits=[-0.15, 0.15],}\\
    & \hspace{1cm} \textit{limit\_units=ppm}, \\
    & Method: "Frequency and phase correction", \\
    & Details: \textit{proc.shift\_to\_reference, ppm\_ref=3.027,}\\
    & \hspace{1cm} \textit{peak\_search=(2.9, 3.1), use\_avg=False}, \\
    & Method: "Phasing", \\
    & Details: \textit{proc.phase\_correct, ppmlim=(2.9, 3.1),}\\
    & \hspace{1cm} \textit{ hlsvd=False, use\_avg=False}, \\
    c. Output measure (e.g. absolute concentration, institutional units, ratio) & Referenced to tCr \\
    d. Quantification references and assumptions, fitting model assumptions & Basis set simulated using MARSS \cite{Landheer2019MagneticRS} \\
    & (metabolites and \acp{mm} from Table~\ref{tab:sim_params}) \\
    & LCModel control:\\
    & \hspace{1cm}\textit{\$LCMODL, nunfil=2048, deltat=0.00025,} \\
    & \hspace{1cm}\textit{hzpppm=127.752504, ppmst=4.2, ppmend=0.5,} \\
    & \hspace{1cm}\textit{dows=F, doecc=F, neach=99, filbas='example.basis',} \\
    & \hspace{1cm}\textit{filraw='example.raw', filps='example.ps',} \\
    & \hspace{1cm}\textit{filcoo='example.coord', lcoord=9, nomit=7,} \\
    & \hspace{1cm}\textit{chomit(1)='MM09', chomit(2)='MM12',} \\
    & \hspace{1cm}\textit{chomit(3)='MM14', chomit(4)='MM17',} \\
    & \hspace{1cm}\textit{chomit(5)='MM20', chomit(6)='-CrCH2',} \\
    & \hspace{1cm}\textit{chomit(7)='CrCH2', namrel='Cr+PCr', \$END} \\
    \midrule
    \textbf{4. Data Quality} & \\
    a. Reported variables (SNR, linewidth (with reference peaks)) & S/N = 16, FWHM = 0.05 - 0.07 ppm (LCModel estimates)\\
    b. Data exclusion criteria & - \\
    c. Quality measures of postprocessing model fitting & - \\
    d. Sample spectrum & See Figure~\ref{fig:invivo_lipids} \\
    \end{tabular*}
\end{table*}

\end{document}